\documentclass[nofootinbib,pra,twocolumn,showpacs,notitlepage,amsmath,amssymb]{revtex4-1} 

\usepackage[colorlinks=true,citecolor=blue,linkcolor=blue,urlcolor=blue]{hyperref}
\usepackage{cleveref}[2012/02/15]
\crefformat{footnote}{#2\footnotemark[#1]#3}

\usepackage{amsmath,amssymb,bm}

\usepackage{tikz}
\usetikzlibrary{external}
\usetikzlibrary{decorations.pathreplacing}
\usetikzlibrary{decorations.text}
\usetikzlibrary{decorations.pathmorphing}
\usetikzlibrary{shapes}

\definecolor{newblue}{RGB}{112,178,255}
\definecolor{neworange}{RGB}{255,204,112}
\definecolor{blue2}{RGB}{120,0,255}
\definecolor{red2}{RGB}{255,0,120}
\definecolor{green2}{RGB}{0,130,130}

\tikzset{snake it/.style={decorate, decoration={snake,segment length=1mm, amplitude=0.5mm}}}

\definecolor{darkred}{RGB}{245,186,183}
\definecolor{lightred}{RGB}{249,217,215}

\usetikzlibrary{arrows,calc,external,shapes.geometric}

\tikzset{
	>=stealth',
	help lines/.style={dashed, thick},
	important line/.style={thick},
	connection/.style={thick, dotted},
}

\tikzstyle{A}=[circle,draw=red!50,fill=red!20,thick]
\tikzstyle{R}=[circle,draw=blue!50,fill=blue!20,thick]
\tikzstyle{U}=[circle,draw=green!50,fill=green!20,thick]
\tikzstyle{V}=[circle,draw=orange!50,fill=orange!20,thick]

\def\bra#1{\mathinner{\langle{#1}|}}
\def\ket#1{\mathinner{|{#1}\rangle}}

\makeatletter
\def\l@subsubsection#1#2{}
\makeatother

\begin{document}
	
	\title{Everything is a quantum Ising model}
	
	\author{Ruben Verresen}
	
	\affiliation{Department of Physics, Harvard University, Cambridge, MA 02138, USA}
	
	\begin{abstract}
		This work shows that any $k$-local Hamiltonian of qubits can be obtained from a 4-state `Ising' model with $k$-local diagonal interactions and a single-site transverse field---giving a new theoretical and experimental handle on quantum matter. In particular, the classical Ising interactions can be determined by replacing each Pauli operator with a $4 \times 4$ diagonal matrix. Subsequently tuning a large transverse field projects out two of the four states, recovering the original qubit model, with qudit generalizations. This leads to striking correspondences, such as the spin-1/2 XY and Heisenberg models arising from the large-field limit of 3-state and 4-state Potts models, respectively. Similarly, the Kitaev honeycomb model emerges from classical interactions which enforce loop states on the honeycomb lattice. These generalized Ising models also display rich physics for smaller fields, including quantum criticality and topological phases of matter.
		This work expands what is experimentally achievable by showing how to realize any quantum spin model using only diagonal interactions and a tuneable field---ingredients found in, e.g., tweezer arrays of Rydberg atoms or polar molecules. More broadly, 4-state spins can also be encoded in the positions of itinerant particles, exemplified by a Bose-Hubbard model realizing the Kitaev honeycomb model---giving an experimental path to its $\mathbb Z_2$ and non-Abelian topological quantum liquids.
	\end{abstract}
	
	\date{\today}
	
	\maketitle
	
	\tableofcontents
	
	\section{Introduction}
	
	The transverse-field Ising model \cite{Katsura62,Pfeuty70} is one of the most elegant many-body quantum systems. While it is simple enough to be exactly solvable in 1D and serve as a workhorse for higher-dimensional numerical simulations, its richness is archetypal for quantum magnetism and universality. Although its original \emph{raison d'\^etre} was to exemplify ordered states of matter, the turn of the millennium showed that adding frustration to the Ising model leads to exotic physics \cite{Moessner00,Moessner01,Priour01,Shokef11,Powalski13,Coester13,Buhrandt14,Roechner16,Sikkenk17,Emonts18,Biswas18,Chamon20,Wu21}. On the experimental front, Rydberg atom tweezer arrays \cite{Endres16,Bernien17,Browaeys_2020,Kaufman2021}---which can now realize 2D \cite{Ebadi20,Scholl20,Semeghini21,Singh21,Bluvstein22} and higher-dimensional geometries \cite{Barredo18,Song21}---are well-described by the quantum Ising model. This has recently led to a resurgence in the study of its phase diagram on various lattices \cite{Fendley04,Samajdar20,Samajdar_2021,Verresen21,Merali21,Slagle21,ORourke22,Slagle22,Giudice22,Rhine22,Kalinowski22,Tarabunga22,Verresen22,Yan23}, sometimes even leading to topological order \cite{Wenbook}.
	
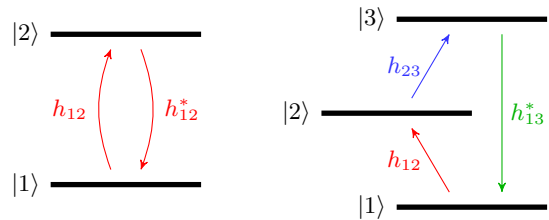
\begin{figure}
		\centering
		\begin{tikzpicture}
		\node at (0,0){
			\begin{tikzpicture}
			\draw[-,line width=2] (0,0) node[left] {$\ket{1}$} -- (2,0) ;
			\draw[-,line width=2] (0,2) node[left] {$\ket{2}$} -- (2,2) ;
			\draw[->,red] (0.8,0.2) to[out=110,in=-110] (0.8,2-0.2);
			\draw[<-,red] (1.2,0.2) to[out=70,in=-70] (1.2,2-0.2);
			\node[right,red] at (1.4,1) {$h_{12}^*$};
			\node[left,red] at (0.6,1) {$h_{12}$};
			\end{tikzpicture}
		};
		\node at (4.1,-0.05){
			\begin{tikzpicture}
			\draw[-,line width=2] (0,-0.25) node[left] {$\ket{1}$} -- (2,-0.25) ;
			\draw[-,line width=2] (-1,1) node[left] {$\ket{2}$} -- (1,1) ;
			\draw[-,line width=2] (0,2.25) node[left] {$\ket{3}$} -- (2,2.25) ;
			\draw[->,red] (0.7,-0.05) -- (0.2,0.8) node[midway,left] {$h_{12}$};
			\draw[->,blue!80] (0.2,1.2) -- (0.7,2.05) node[midway,left] {$h_{23}$};
			\draw[->,green!70!black] (1.4,2.05) -- (1.4,-0.05) node[midway,right] {$h_{13}^*$};
			\end{tikzpicture}
		};
	\end{tikzpicture}
	\caption{\textbf{The importance of going beyond 2-state Ising models.} A single-site field $H_\textrm{field} = -\sum_{i,j} h_{ij} \ket{i} \bra{j}$ can have non-trivial transition cycles for a $q$-state Ising model only if $q > 2$. E.g., cycling through $\ket{1} \to \ket{2} \to \ket{3} \to \ket{1}$ has gauge-invarant amplitude $h_{12} h_{23} h_{13}^*$ (i.e., this is unaffected by redefining basis states). If $h_{12} h_{23} h_{13}^*$ has a nonzero phase factor, the Ising model is not stoquastic (it has a so-called `sign-problem'), allowing for the emergence of rich physics. Indeed, this work shows that higher-state Ising models can replicate all phenomena in quantum magnetism.}
	\label{fig:threestatelevels}
\end{figure}

	This plethora of phenomenology raises the question: \emph{what can the quantum Ising model (not) do?} A main constraint is that it is sign-problem-free\footnote{To wit, this means that there exists a basis where off-diagonal operators have only negative coefficients.}; while this is an advantage for simulating the model with quantum Monte Carlo \cite{Sandvikreview}, it implies that many (indeed, most) phases of matter cannot arise as its ground state \cite{Hastings16,Ringel17,Smith20,Golan20}. One underappreciated way of removing this restrictive property is by simply going beyond qubits: e.g., a magnetic field on a \emph{three}-level system allows for cycles with nonzero phase factors\footnote{The author thanks Ashvin Vishwanath for an illuminating discussion on this point.}; see Fig.~\ref{fig:threestatelevels}. In this work, we show that this minimal change to the quantum Ising model encompasses all other bosonic models with finite-dimensional on-site Hilbert spaces! Indeed, any quantum spin model is shown to arise as the large-field limit of such a generalized Ising model---generically on 4-state spins, although sometimes three states suffice.
	
	The relationship we establish between an arbitrary quantum spin model and an Ising model is quite direct and user-friendly. Let us first observe that without loss of generality, we can restrict to spin-1/2 magnets.
	Indeed, a qudit can always be identified with (a subspace of) multiple qubits, which preserves spatial locality of interactions whilst potentially introducing multi-body terms\footnote{Alternatively, such multi-body interactions could be avoided by instead obtaining a $p$-local $q$-state quantum model from a $p$-local $q^2$-state transverse field Ising model; see Sec.~\ref{sec:outlook}.}. For an arbitrary spin-1/2 model, the idea is then to simply replace the Pauli operators $\sigma^\alpha$ by certain diagonal operators $\mathcal Z^\alpha$. While this `unquantization' gives a classical model which is seemingly unrelated to the original quantum model, we show that the spin-1/2 model re-emerges upon adding a large transverse field. This simple prescription distinguishes it from other works on universal models \cite{vandenNest08,DelasCuevas09,DelasCuevas10,DelasCuevas16,Cubitt18,Kohler19}, which typically involve rather nonlocal encodings and deep proofs based on complexity theory. The cost we pay for such a direct relationship is that if we wish to obtain a $k$-local qubit model, the corresponding Ising model is also $k$-local; we thus do \emph{not} reduce \emph{multi-body} quantum spin models to a \emph{two-body} Ising model.
	
	The consequences are at least twofold. Firstly, it establishes a conceptual connection between quantum and classical interactions, leading to new rich models ripe for exploration. After stating our general results (Sec.~\ref{sec:general}), we showcase this for various paradigmatic quantum magnets. In the case of the XY (Sec.~\ref{subsec:xy}) and Heisenberg (Sec.~\ref{subsec:heisenberg}) magnets, we find that their phenomenology can persist to the regime where the classical interactions are dominant. For instance, staggering a 4-state Potts chain can stabilize a symmetry-protected topological (SPT) phase \cite{Gu09,pollmann_entanglement_2010,pollmann_symmetry_2012,Fidkowski_2011,Schuch11,chen_complete_2011,Senthil_2015}, even for arbitrarily small transverse fields. The present work also further explores a connection between the Kitaev honeycomb model and the 4-state Ising interactions which realize a dimer liquid \cite{RK,Sachdev92,Sachdev99,Moessner_2001,Moessner01b,Misguich02,Fradkin_2013}, which the author recently established in collaboration with Ashvin Vishwanath \cite{Verresen22}. We show that this provides a solvable model for studying the interpolation between symmetry-enriched $\mathbb Z_2$ spin liquids \cite{FradkinShenker,Read91,Wen,Sachdev99,Essin13,Mesaros13,Hung13,Hung13b,Chen15,Lu16,Lee18,Barkeshli_2019} where we find an intervening non-Abelian phase (Sec.~\ref{subsec:kitaev}). A second consequence is that these generalized Ising models provide an alternative and minimal framework for experimentally realizing various models of interest, which we illustrate in Sec.~\ref{sec:exp} for Rydberg and dipolar systems.
	
\section{From quantum spin model to Ising model and back \label{sec:general}}

In this section we present the general results. We refer the reader who prefers an example-based approach to Sec.~\ref{sec:examples}, which contains several fleshed-out case studies.

\subsection{General case: 4-state model \label{subsec:4}}

We consider an arbitrary lattice $\Lambda$ of spin-1/2's, where each site has Pauli operators satisfying the algebra $[\sigma^z_j,\sigma^+_j] = \sigma^+_j$ and $[\sigma^+_j, \sigma^-_j] = \sigma^z_j$.
An arbitrary Hamiltonian can be written as a function which is linear in the Pauli operators for each site:
\begin{equation}
	H = f\left( \left\{ \sigma^+_j,\sigma^-_j,\sigma^z_j \right\}_{j \in \Lambda} \right). \label{eq:Hgeneral}
\end{equation}
(E.g., $f(\{a_j,b_j,c_j\}) = \sum_{\langle i,j\rangle} \left( a_i b_j + b_i a_j \right)$ gives the XY model; see Sec.~\ref{sec:examples} for more examples.) We associate to $H$ a generalized 4-state Ising model $\tilde H$ (on the same lattice) by replacing the Pauli operators by diagonal ones:
\begin{equation}
	\tilde H = f \left( \left\{ \sqrt{\frac{3}{2}} e^{i\phi} \mathcal Z_j,\sqrt{\frac{3}{2}}  e^{-i\phi}\mathcal Z_j^\dagger,\sqrt{3} \mathcal Z_j^2 \right\}\right) + \lambda \sum_j \mathcal X_j \label{eq:H4state}
\end{equation}
where for every site we have the 4-state matrices
\begin{equation}
	\mathcal Z =\left( \begin{array}{cccc} 1 & 0 & 0 & 0 \\0 & i & 0 & 0\\ 0 & 0 & -1 & 0 \\ 0& 0& 0& -i \end{array} \right), \; 
	\mathcal X = \left( \begin{array}{rrrr} 0 & 1 & 1 & 1 \\
		1 & 0 & -i & i \\
		1 & i & 0 & -i \\
		1 & -i & i & 0
	\end{array} \right). \label{eq:4stateX}
\end{equation}
One can show that the large-field limit projects each site into the low-energy doublet of $\mathcal X$, thereby recovering $H$:
\begin{equation}
	\tilde H \xrightarrow{\lambda \to \infty} P\tilde H P = H.
	\label{eq:general}
\end{equation}

(The phase $e^{i\phi}$ in Eq.~\eqref{eq:H4state} can be chosen freely. We note that hermiticity of $H$ ensures that $\tilde H$ is hermitian. Moreover, if $H$ is real, then $\tilde H$ with $\phi=0$ has an anti-unitary symmetry\footnote{Indeed, $\tiny T =  \left( \begin{array}{cccc}
	1 & 0 & 0 & 0 \\
	0 & 0 & 0 & 1 \\
	0 & 0 & 1 & 0 \\
	0 & 1 & 0 & 0 \end{array} \right) K$ commutes with $\mathcal Z$ and $\mathcal X$.}. Nevertheless, it can sometimes be useful to take $\phi\neq 0$; see Sec.~\ref{subsec:kitaev}.)

\emph{Proof of Eq.~\eqref{eq:general}:} A straightforward computation shows that the spectrum of $\mathcal X$ is $\{\sqrt{3},\sqrt{3},-\sqrt{3},-\sqrt{3}\}$. The $\lambda \to \infty$ limit energetically enforces $\mathcal X = -\sqrt{3}$, leaving a two-dimensional Hilbert space per site. Let $P = \frac{\sqrt{3} \; \mathbb I_4 - \mathcal X}{2\sqrt{3}}$ denote the projector onto this subspace and define
\begin{equation}
	\tilde \sigma^+_j = \sqrt{\frac{3}{2}} e^{i\phi} (P\mathcal ZP)_j \quad \textrm{and} \quad
	\tilde \sigma^z_j = \sqrt{3} (P\mathcal Z^2P)_j.
\end{equation}
Then at leading order in perturbation theory, the effective Hamiltonian is $
\lim_{\lambda \to \infty} 
\tilde H = f \left( \left\{ \tilde \sigma^+_j, \tilde \sigma^-_j, \tilde \sigma^z_j \right\} \right)$, where $\tilde \sigma^-_j = \left( \tilde \sigma^+_j \right)^\dagger$.
Crucially, one can check that these matrices respect the Pauli algebra, e.g., $[\tilde \sigma^z,\tilde \sigma^+] = 2\tilde \sigma^+$ and $[\tilde \sigma^+, \tilde \sigma^-] = \tilde \sigma^z$. QED.

Note that if the quantum spin model is written as $H = g(\{ \sigma^x_j , \sigma^y_j , \sigma^z_j \} )$, then the corresponding Ising model is obtained by substituting $\sigma^\alpha \to \sqrt{3} \mathcal Z^\alpha$ where
\begin{equation}
	\mathcal Z^x_j = \frac{e^{i\phi} \mathcal Z_j + e^{-i \phi} \mathcal Z^\dagger_j}{\sqrt{2}}, \;
	\mathcal Z^y_j = \frac{e^{i\phi} \mathcal Z_j - e^{-i \phi} \mathcal Z^\dagger_j}{\sqrt{2}i}, \;
	\mathcal Z^z_j = \mathcal Z_j^2. \label{eq:ZxZyZz}
\end{equation}

\subsection{Intuition \label{subsec:intuition}}

While the above proof is rigorous, it is ad hoc. Here we present an alternative approach, which can provide some intuition and can perhaps serve as a basis for future generalizations.

Let $\sigma^{\alpha =x,y,z}$ and $\tau^{\alpha =x,y,z}$ denote Pauli matrices. If we define $\mathcal Z^\alpha =- \sigma^{\alpha} \otimes  \tau^{\alpha}$, then these three $4 \times 4$ matrices clearly mutually commute, and hence they can be simultaneously diagonalized. (In fact, these can be related to the three diagonal matrices in Eq.~\eqref{eq:ZxZyZz}.) However, if we impose a large energetic term $\lambda \sum_\alpha \mathbb I \otimes \tau^\alpha$ (with $\lambda \to +\infty$), then at low energies we pin $\tau := \frac{\tau^x + \tau^y + \tau^z}{\sqrt{3}} = -1$. In this limit, we can substitute $\tau^\alpha \to \tau/\sqrt{3} = -1/\sqrt{3}$. Hence, as $\lambda \to \infty$, we have that ${\mathcal Z}^\alpha \to \sigma^\alpha/\sqrt{3}$, i.e., the diagonal matrices reduce to Pauli matrices.

To summarize the basic idea which could be more generally applicable: one can pair up non-commuting matrices into commuting (and hence diagonal) ones, after which single-site energetics can freeze out one of the two to recover the original non-commuting algebra.

\subsection{Special case: 3-state model \label{subsec:3}}

In certain cases (e.g., the XY model) the interactions do not use all three Pauli components. Here we show that such models can be obtained from a generalized Ising model on a 3-state spin, rather than the more general 4-state spin above.

In particular, suppose we have
\begin{equation}
	H = f\left( \left\{ \sigma^+_j,\sigma^-_j \right\} \right) + \sum_j h_j \sigma^z_j. \label{eq:Hspecial}
\end{equation}
We associate to this a 3-state Ising model:
\begin{equation}
	\tilde H = H\left( \left\{ \mathcal Z_j, \mathcal Z_j^\dagger \right\} \right) + \sum_j \left[ \left( \lambda + i \frac{h_j}{\sqrt{3}} \right) \mathcal X_j + h.c. \right] \label{eq:H3state}
\end{equation}
with
\begin{equation}
	\mathcal Z = e^{i\phi} \left( \begin{array}{ccc}
		1 & 0 & 0 \\
		0 & \omega & 0 \\
		0 & 0 & \bar \omega \end{array} \right) \textrm{ and } \mathcal X = \left( \begin{array}{ccc}
		0 & 1 & 0 \\
		0 & 0 & 1 \\
		1 & 0 & 0 \end{array} \right) \label{eq:3statematrices}
\end{equation}
where $e^{i\phi}$ is arbitrary and $\omega= e^{\frac{2\pi i}{3}}$. It can be shown that
\begin{equation}
	\tilde H \xrightarrow{\lambda \to \infty} P\tilde H P = H. \label{eq:3thm}
\end{equation}
This can be proven similarly to the general case above, now using the projector $P = \frac{2-\mathcal X - \mathcal X^\dagger}{3}$ and defining $\tilde \sigma^+_j = \left(P \mathcal Z P \right)_j$ and $\tilde \sigma^z_j=i\frac{\left(P\mathcal X P\right)_j}{\sqrt{3}} + h.c.$.

If $h_j=0$, the third Pauli component never appears, and then $\tilde H$ is manifestly real. Indeed, the diagonal term is always real due to hermiticity. 

\section{Examples \label{sec:examples}}

Here we illustrate the above general results for a few archetypal models of quantum magnetism.

\subsection{Spin-1/2 XY model \label{subsec:xy}}
Let us first consider the spin-1/2 XY model, $H = J \sum_{\langle i,j\rangle} \left(\sigma^+_i \sigma^-_j + h.c. \right)$. In this case, we can use the special result in Sec.~\ref{subsec:3}, saying that it arises as the large-field limit of a 3-state model with Hamiltonian
\begin{align}
	\tilde H 
	&= J \sum_{\langle i,j\rangle} \left( \mathcal Z_i \mathcal Z_j^\dagger + h.c. \right) + \lambda \sum_j \left(\mathcal X_j + \mathcal X_j^\dagger \right) \\
	&= 3J \sum_{\langle i,j\rangle} \delta_{ij} + \lambda \sum_j \left( \begin{array}{ccc} 0 & 1 & 1 \\ 1 & 0 & 1 \\ 1 & 1 & 0 \end{array} \right)_j  \label{eq:3statePotts} + \textrm{const.}
\end{align}
The classical interactions are those of a Potts model, which has an $S_3$ symmetry permuting the three states.
We see that for large $\lambda$, the single-site field projects out the state $\frac{1}{\sqrt{3}} \left( \ket{0} + \ket{1} + \ket{2} \right)$, giving an effective qubit model. More precisely, Eq.~\eqref{eq:3thm} says that if $\lambda \to + \infty$ we obtain the spin-1/2 XY model, where $S_3 \cong \mathbb Z_3 \rtimes \mathbb Z_2$ is enhanced to $O(2) \cong U(1) \rtimes \mathbb Z_2$ symmetry.

To gain some more insight into the physics of this model, let us discuss the 1D case. The spin-1/2 XY chain is exactly solvable \cite{Lieb61} and is described by a conformal field theory (CFT) at low energies \cite{ginsparg_applied_1988}, namely, the compact boson CFT or Luttinger liquid with parameter $K=1$ \cite{AffleckLesHouches}. The scaling dimension of the charge-3 operator is $\frac{9}{4K} = \frac{9}{4}$, which is larger than the spacetime dimension and hence irrelevant. This shows that the CFT is robust even away from the $\lambda \to +\infty$ limit. In fact, one can perform a more detailed analysis involving perturbation theory (see Appendix~\ref{app:3state}) which suggests that for $J>0$, the gapless phase and its emergent $U(1)$ symmetry is stable for all field strengths $\lambda >0$. This agrees with a recent numerical study \cite{Dai17}. Hence, one can interpret the robust gapless phase of the antiferromagnetic 3-state Potts chain as essentially realizing the spin-1/2 XY chain.

\subsection{Spin-1/2 XXZ and Heisenberg model \label{subsec:heisenberg}}

Another paradigmatic magnet is the spin-1/2 XXZ model,
$H(\Delta) = \frac{J}{4} \sum_{\langle i ,j\rangle} \left( \sigma^x_i \sigma^x_j + \sigma^y_i \sigma^y_j + \Delta \sigma^z_i \sigma^z_j\right)$.
According to Sec.~\ref{subsec:4}, this arises as the large-field limit of the following 4-state Ising model:
\begin{equation}
	\tilde H(\Delta) = \frac{3J}{4} \sum_{\langle i,j\rangle} \left( \mathcal Z_i \mathcal Z_j^\dagger + \mathcal Z_i^\dagger \mathcal Z_j + \Delta \mathcal Z_i^2 \mathcal Z_i^2 \right) + \lambda \sum_{j} \mathcal X_j, \label{eq:4stateXXZ}
\end{equation}
where $\mathcal Z$ and $\mathcal X$ are defined in Eq.~\eqref{eq:4stateX}. The particular case of the Heisenberg model corresponds to:
\begin{equation}
	\tilde H(\Delta = 1) = 3J \sum_{\langle i,j\rangle} \delta_{ij} +\lambda \sum_j \mathcal X_j. \label{eq:4stateHeisenberg}
\end{equation}
We recognize this as the 4-state Potts model with an unusual complex-valued field \eqref{eq:4stateX}. The latter is unavoidable if one wants to recover such spin-1/2 models in the large-field limit: the Pauli algebra $[\sigma^\alpha,\sigma^\beta]=2i\varepsilon_{\alpha \beta \gamma} \sigma^\gamma$ involves complex numbers, whereas diagonal hermitian matrices (such as those in Eq.~\eqref{eq:ZxZyZz}) are real. To gain some general insight into this novel Potts model, let us discuss its symmetries and related anomalies.

\emph{Symmetries.} The 4-state model in Eq.~\eqref{eq:4stateXXZ} has a $\mathbb Z_2 \times \mathbb Z_2$ symmetry generated by $\prod_j U_j^{\alpha=y,z}$ where
\begin{equation}
	U^y = {\footnotesize \left( \begin{array}{cccc}
			0 & 1 & 0 & 0 \\
			1 & 0 & 0 & 0 \\
			0 & 0 & 0 & i \\ 
			0 & 0 & -i & 0 \end{array} \right)}
	\; \textrm{ and } \;
	U^z = {\footnotesize \left( \begin{array}{cccc}
			0 & 0 & 1 & 0 \\
			0 & 0 & 0 & -i \\
			1 & 0 & 0 & 0 \\ 
			0 & i & 0 & 0 \end{array} \right)}. \label{eq:Z2Z2}
\end{equation}
In the limit $\lambda \to +\infty$, this corresponds to the $\mathbb Z_2 \times \mathbb Z_2$ symmetry of the spin-1/2 XXZ model. The isotropic case \eqref{eq:4stateHeisenberg} has an enhanced symmetry: Potts interactions have a natural $S_4$ symmetry, which is broken down to $A_4$
by the complex field, as described in Appendix~\ref{app:4state}.
There we also define an anti-unitary symmetry such that we have $A_4 \rtimes \mathbb Z_2^T \cong S_4^T$. When $\lambda \to + \infty$, the discrete $A_4$ symmetry corresponds to the tetrahedral subgroup of the $SO(3)$ symmetry of the Heisenberg model.

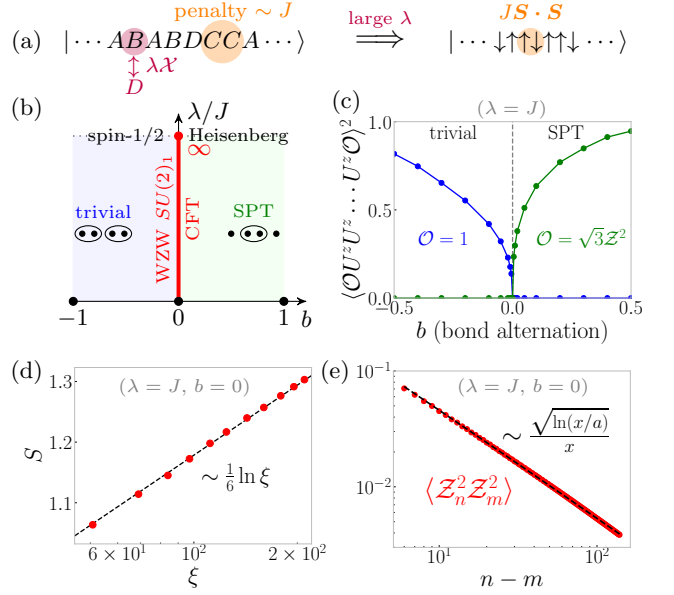
\begin{figure}
	\centering
	\begin{tikzpicture}
	\filldraw[orange,opacity=0.3] (0.5,0) circle (8pt);
	\filldraw[purple,opacity=0.3] (-0.67,0) circle (5pt);
	\node at (0,0) {$|\cdots ABABDCCA \cdots\rangle$};
	\node[orange] at (0.65,0.4) {\footnotesize penalty $\sim J$};
	\node[purple] at (-0.42,-0.33) {\footnotesize $\updownarrow \lambda \mathcal X$};
	\node[purple] at (-0.67,-0.6) {\footnotesize $D$};
	\node at (-2.1,0.) {(a)};
	\node at (2.6,0) {\large $\Longrightarrow$};
	\node[purple] at (2.6,0.3) {\scriptsize large $\lambda$};
	\filldraw[orange,opacity=0.3] (4.6,0) circle (5pt);
	\node at (4.7,0) {$|\cdots \downarrow \uparrow \uparrow \downarrow \uparrow \uparrow \downarrow \cdots\rangle$};
	\node[orange] at (4.6,0.4) {\footnotesize $J \boldsymbol{S \cdot S}$};
	\node at (0,-2.5+0.2) {
		\begin{tikzpicture}
		\filldraw[blue,opacity=0.05] (-1.4,0) -- (-1.4,2.2) -- (0,2.2) -- (0,0);
		\filldraw[green,opacity=0.05] (1.4,0) -- (1.4,2.2) -- (0,2.2) -- (0,0);
		\draw[dotted] (-1.4,2.2) -- (1.4,2.2);
		\draw[->] (-1.4,0) -- (1.7,0) node[below] {$b$};
		\draw[->] (0,0) -- (0,2.5) node[right] {$\lambda/J$};
		\draw[-,red,line width=1.5] (0,0) -- (0,2.2) node[above,midway,rotate=90,yshift=-2] {\scriptsize WZW $SU(2)_1$} node[below,midway,rotate=90,yshift=1] {\scriptsize CFT};
		\filldraw (-1.4,0) circle (1.5pt) node[below] {$-1$};
		\filldraw (0,0) circle (1.5pt) node[below] {$0$};
		\filldraw (1.4,0) circle (1.5pt) node[below] {$1$};
		\filldraw[red] (0,2.2) circle (1.5pt) node[below right] {$\infty$};
		\node at (-0.7,2.2) {\scriptsize spin-1/2};
		\node at (0.8,2.2) {\scriptsize Heisenberg};
		\node[blue] at (-1,1.2) {\scriptsize trivial};
		\node[black!50!green] at (1,1.2) {\scriptsize SPT};
		\draw (-1.2,0.9) ellipse (5pt and 3pt);
		\draw (-0.8,0.9) ellipse (5pt and 3pt);
		\filldraw (-1.3+0.02,0.9) circle (1pt);
		\filldraw (-1.1-0.02,0.9) circle (1pt);
		\filldraw (-0.9+0.02,0.9) circle (1pt);
		\filldraw (-0.7-0.02,0.9) circle (1pt);
		\draw (1,0.9) ellipse (5pt and 3pt);
		\filldraw (1.3,0.9) circle (1pt);
		\filldraw (1.1-0.02,0.9) circle (1pt);
		\filldraw (0.9+0.02,0.9) circle (1pt);
		\filldraw (0.7,0.9) circle (1pt);
		\end{tikzpicture}};
	\node at (-2.1,-0.9) {(b)};
	\node at (4.08,-2.5) {\includegraphics[scale=0.25]{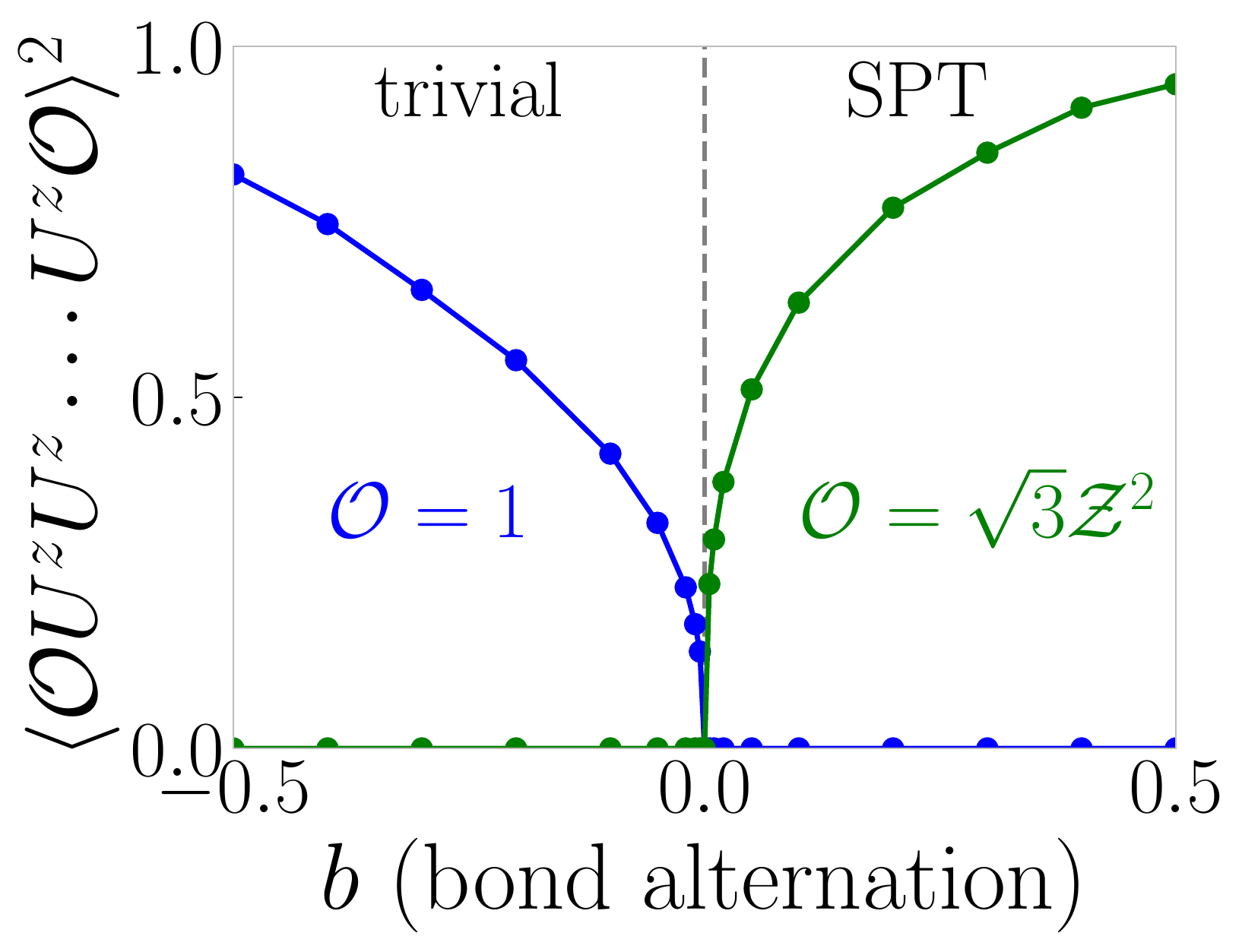}};
	\node at (2.15,-0.8) {(c)};
	\node at (-0.12,-2.5-3.3) {\includegraphics[scale=0.25]{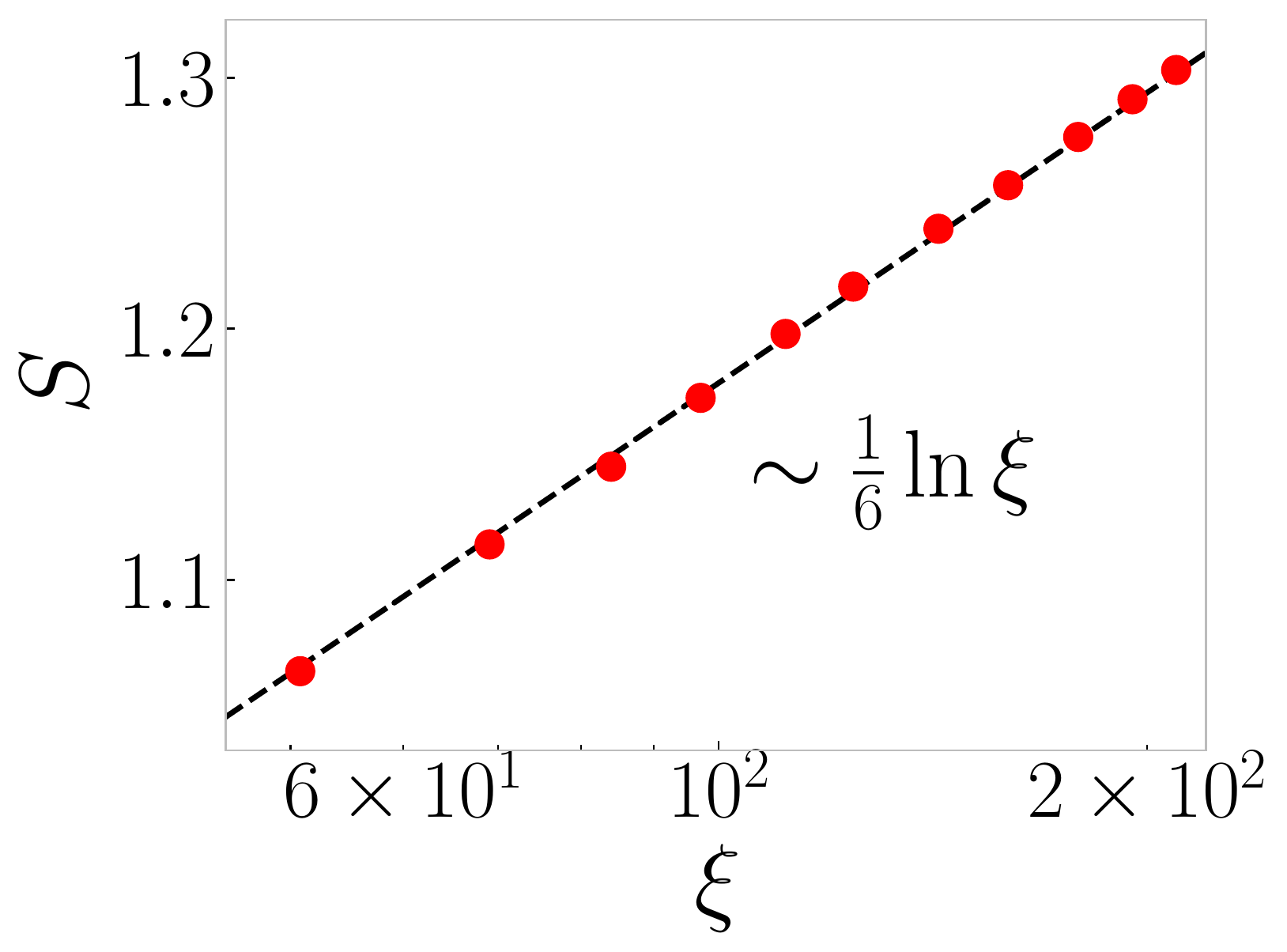}};
	\node at (4.1,-2.5-3.3) {\includegraphics[scale=0.25]{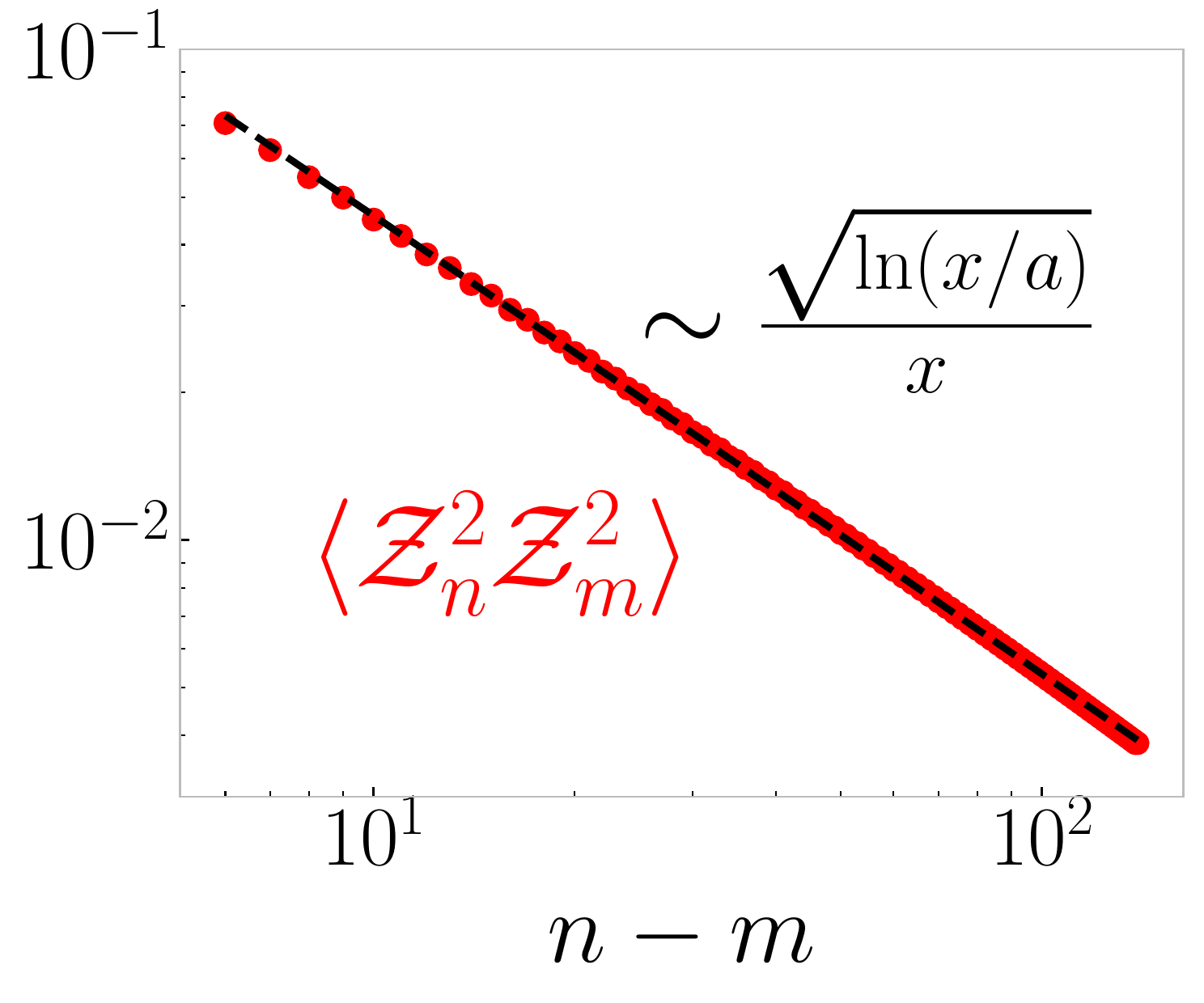}};
	\node at (-2.1,-1.1-3.3) {(d)};
	\node at (2.,-1.1-3.3) {(e)};
	\node[color=gray] at (0,-4.6) {\scriptsize ($\lambda=J$, $b=0$)};
	\node[color=gray] at (4.5,-4.6) {\scriptsize ($\lambda=J$, $b=0$)};
	\node[color=gray] at (4.35,-0.9) {\scriptsize ($\lambda=J$)};
\end{tikzpicture}
\caption{\textbf{Spin-1/2 Heisenberg chain and Haldane SPT phase from an antiferromagnetic Potts model.} (a) We consider a 4-state Potts model \eqref{eq:4stateHeisenberg} where neighbors experience an energy cost $J>0$ if they are in the same state; a complex field \eqref{eq:4stateX} generates resonances. For large field, we obtain the spin-1/2 Heisenberg chain, described by the Wess-Zumino-Witten (WZW) $SU(2)_1$ conformal field theory. (b) We numerically find that this criticality persists to all values of the field, stabilized by the $A_4 \subset SO(3)$ symmetry of the Potts chain. Moreover, staggering the Potts interactions leads to an SPT phase protected by $A_4$. (c) The (non)trivial SPT phases can be detected by string order parameters; shown for $\lambda=J$. (d) At the translation-invariant point, entanglement scaling is consistent with central charge $c=1$. (e) Similarly we confirm the scaling dimension $\Delta=1/2$ of the spin operator in the WZW CFT (with logarithmic corrections \cite{Giamarchi89,Singh89,Nomura93}).}
\label{fig:Potts}
\end{figure}

\emph{Anomalies.} In spin-1/2 models, spin-rotation acts projectively on a single spin. This continues to hold for our generalized 4-state Ising models. For instance, $U^y$ and $U^z$ defined in Eq.~\eqref{eq:Z2Z2} anticommute on a single site. Similarly, Eq.~\eqref{eq:4stateHeisenberg} is symmetric under a projective representation of $A_4$, distinguishing it from the usual 4-state Potts model with a real field. Such symmetries can give powerful constraints on the phase diagram. For instance, when combined with certain spatial symmetries, projective symmetry actions are `anomalous' \cite{Lieb61,Oshikawa97,Yamanaka97,Oshikawa00,Misguich02b,Hastings04,Tasaki04,Hastings05,Nachtergaele07,Parameswaran13,Nomura15,Watanabe15,Cheng16,Po17,Cho17,Metlitski18,Jian_2018,Yang_2018,Takahashi20,Wang21}, meaning that the ground state cannot be a trivial symmetric phase of matter. One rich example is a square lattice of spin-1/2's with $C_4$ rotation symmetry: this anomaly stabilizes a direct non-Landau transition between a valence bond solid and N\'eel state---a `deconfined quantum critical point' (DQCP) described by a putative $SO(5)$ field theory \cite{Senthil04,Senthil04b,Sandvik07b,Sandvik10,Nahum15,Wang_2017,Li19,Serna19}. Such phenomena can thus be explored in 4-state Ising models, even away from the large-field limit. In fact, even though finite $\lambda$ reduces $SO(3)$ to $A_4$, this symmetry stabilizes the DQCP \cite{Wang_2017,Metlitski18,Tantivasadakarn21}.

\emph{1D case---criticality.} Let us study the physics of the 4-state Potts chain \eqref{eq:4stateHeisenberg}, as shown in Fig.~\ref{fig:Potts}(a). In the limit $\lambda \to +\infty$, this is the spin-1/2 Heisenberg chain with a Lieb-Schultz-Mattis anomaly \cite{Lieb61} due to spin-rotation and translation symmetry. The ground state is a Luttinger liquid at the $SU(2)$-symmetric $K=1/2$ \cite{WZW,AffleckLesHouches}. The $A_4$ symmetry stabilizes this anomaly and criticality to finite $\lambda$. Remarkably, we find that this holds for \emph{all} $\lambda>0$ via numerical density matrix renormalization group (DMRG) \cite{White92,White93} simulations using the TeNPy library \cite{Hauschild18}. A moderate bond dimension $\chi = 150$ was sufficient to obtain converged results for Fig.~\ref{fig:Potts}. We observe the expected central charge based on entanglement scaling \cite{Calabrese04,Pollmann09}, and even the spin-spin correlations associated to $K=1/2$ \cite{Giamarchi89,Singh89,Nomura93}. The microscopic $A_4$ symmetry thus gives a low-energy emergent $SU(2)$ symmetry.

\emph{1D case---Haldane SPT.} The aforementioned anomaly is related to how staggering the Potts interactions, $\sum_n (1+b(-1)^n ) \delta_{n,n+1}$, gives rise to two distinct symmetry-protected topological (SPT) phases protected by $A_4$. (Indeed, one can interpret single-site translation as an SPT-entangler.) Similar to the Su-Schrieffer-Heeger chain \cite{SSH}, it is conventional to fix a two-site unit cell \cite{deLeseleuc19,Sompet21} say $(2n-1,2n)$, after which $b<0$ ($b>0$) is the trivial (topological) phase. This is evidenced by the trivial string order parameter $\langle \prod_{i<m<j} U^z_{2m-1} U^z_{2m} \rangle$ having long-range order only for $b<0$; for the SPT phase we need to include an endpoint operator which is odd under $U^y$ \cite{Pollmann12}, such as $\mathcal Z^2$; see Fig.~\ref{fig:Potts}(c). We have also confirmed that for $b>0$, the entanglement spectrum is twofold degenerate (not shown), which persists even upon explicitly breaking bond-centered inversion symmetry (which is also able to protect the phase \cite{pollmann_entanglement_2010,pollmann_symmetry_2012}). In the large-field limit, this reduces to the bond-alternating spin-1/2 Heisenberg chain, which moreover connects to the spin-1 Heisenberg chain \cite{Haldane83,Haldane83b,affleck1988,pollmann_symmetry_2012} upon making the intra-unit-cell couplings ferromagnetic \cite{Hida92,White96}.

\subsection{Kitaev honeycomb model \label{subsec:kitaev}}

\begin{figure}
\centering
\includegraphics[scale=0.6]{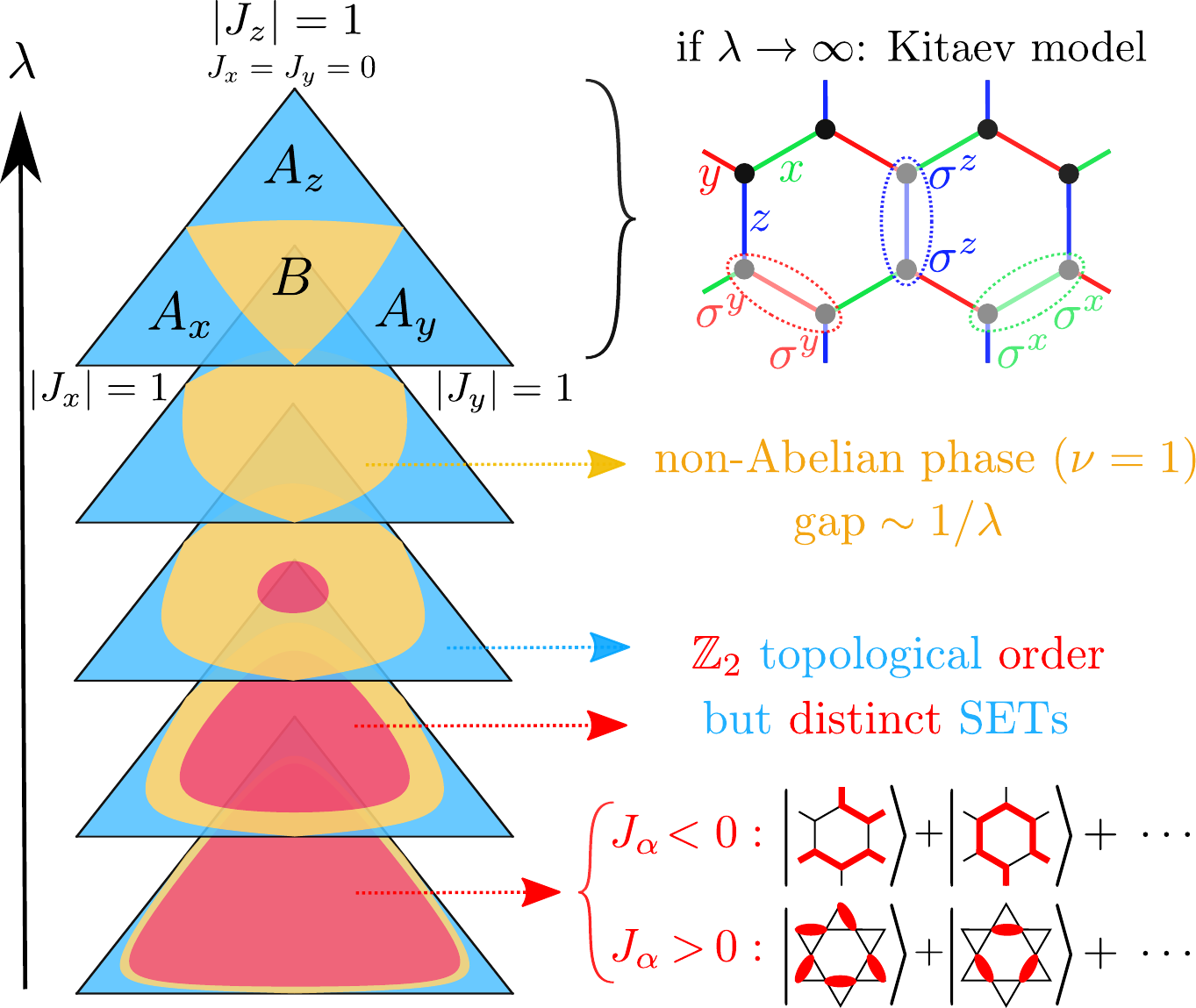}
\caption{\textbf{Kitaev honeycomb model from a generalized transverse-field Ising model.} The large-field limit $\lambda \to \infty$ of Eq.~\eqref{eq:Hisingkitaev} leads to the spin-1/2 Kitaev model. The isotropic case $|J_x| =|J_y| = |J_z|$ was recently studied in Ref.~\onlinecite{Verresen22}; here we study the full phase diagram, plotted in barycentric coordinates $|J_x| +|J_y| + |J_z|= 1$ for different values of $\lambda \approx 0.23, 0.40, 0.56, 0.87,2.31$. The gapless (`$B$') phase of the Kitaev model \cite{Kitaev06} is gapped out into a non-Abelian chiral phase shown in yellow (see the main text for a different Ising model where the Majorana cones are robust). The blue regions are the gapped $\mathbb Z_2$ spin liquid phases of the Kitaev model. For small fields (red), the antiferromagnetic model gives an emergent dimer liquid on the kagom\'e lattice \cite{Verresen22} whereas the ferromagnetic case gives a honeycomb loop model (akin to the toric code state \cite{Kitaev03}); both are distinct from the $A_\alpha$ phases in the presence of translation symmetry. }
\label{fig:kitaev}
\end{figure}

Let us now consider the spin-1/2 Kitaev honeycomb model \cite{Kitaev06}: $H = \sum_{\alpha=x,y,z} J_\alpha  \sum_{\langle i,j \rangle_\alpha} \sigma^\alpha_i \sigma^\alpha_j$, with the bond-dependent couplings shown in Fig.~\ref{fig:kitaev}. According to Sec.~\ref{subsec:4}, this arises as the large-field limit of the following 4-state Ising model:
\begin{equation}
\tilde H = 3\sum_{\alpha=x,y,z}J_\alpha  \sum_{\langle i,j \rangle_\alpha} \mathcal Z^\alpha_i \mathcal Z^\alpha_j + \lambda \sum_j \mathcal X_j \label{eq:Hisingkitaev}
\end{equation}
where the diagonal operator $\mathcal Z^\alpha$ is defined in Eq.~\eqref{eq:ZxZyZz}, and $\mathcal X$ in Eq.~\eqref{eq:4stateX}. We briefly recall the phase diagram for $\lambda \to +\infty$, i.e., that of the Kitaev model: there is a gapless $\mathbb Z_2$ spin liquid (labeled $B$ in Fig.~\ref{fig:kitaev}) within the triangle inequality $|J_z| < |J_x| + |J_y|$ (and permutations thereof), whereas outside of these bounds there are three $\mathbb Z_2$ spin liquids (labeled $A_{\alpha=x,y,z}$) which are distinct in the presence of translation symmetry \cite{Kitaev06}.

What is the fate of this celebrated phase diagram if we make $\lambda$ finite? This depends on our choice of $\phi$ in Eq.~\eqref{eq:ZxZyZz}. (In our previous examples this did not arise since $\mathcal Z_i$ always appeared together with $\mathcal Z^\dagger_j$.) In Sec.~\ref{subsec:4} we saw that if we choose $\phi = 0$, then $\tilde H$ has an anti-unitary symmetry; this is sufficient to stabilize the gapless $\mathbb Z_2$ spin liquid \cite{Kitaev06}. Hence, for one choice of generalized Ising model, the $B$ phase will be robust for some window $\lambda \geq \lambda_c$; we leave the study of this model to future work. Here, we set $\phi = \frac{\pi}{4}$: this breaks time-reversal symmetry for finite $\lambda$ and will hence gap out the $B$ phase into a non-Abelian phase \cite{Kitaev06}, but more importantly, it turns out to preserve the exact solubility of the Kitaev model for any $\lambda$; see Appendix~\ref{app:kitaev}. In this case, one can show that the signs of $J_\alpha$ can be unitarily toggled.

Having set $\phi = \frac{\pi}{4}$, let us first briefly discuss the classical limit, i.e., Eq.~\eqref{eq:Hisingkitaev} with $\lambda=0$. There is an extensive ground state degeneracy $\sim \sqrt{2}^N$, where $N$ is the total number of sites. Although the signs of $J_\alpha$ can be unitarily toggled, the interpretation of this classical degeneracy depends on the choice of sign. In the ferromagnetic case $J_\alpha<0$, it is convenient to label the basis states of each 4-state spin as $\big\{
\!\!
\raisebox{-2pt}{
\begin{tikzpicture}[scale=0.4]
\draw[-] (0,0) -- (0.5,0.2887) -- (0.5,0.866);
\draw[-] (0.5,0.2887) -- (1,0);
\end{tikzpicture}
}
\!\!,\!\!
\raisebox{-2pt}{
\begin{tikzpicture}[scale=0.4]
\draw[-] (0,0) -- (0.5,0.2887) -- (0.5,0.866);
\draw[-] (0.5,0.2887) -- (1,0);
\draw[-,red,line width=2] (0,0) -- (0.5,0.2887) -- (0.5,0.866);
\end{tikzpicture}
}
\!\!,\!\!
\raisebox{-2pt}{
\begin{tikzpicture}[scale=0.4]
\draw[-] (0,0) -- (0.5,0.2887) -- (0.5,0.866);
\draw[-] (0.5,0.2887) -- (1,0);
\draw[-,red,line width=2] (0,0) -- (0.5,0.2887) -- (1,0);
\end{tikzpicture}
}
\!\!,\!\!
\raisebox{-2pt}{
\begin{tikzpicture}[scale=0.4]
\draw[-] (0,0) -- (0.5,0.2887) -- (0.5,0.866);
\draw[-] (0.5,0.2887) -- (1,0);
\draw[-,red,line width=2] (0.5,0.866) -- (0.5,0.2887) -- (1,0);
\end{tikzpicture}
}
\!\!
\big\}$ on one sublattice of the honeycomb lattice and as 
$\big\{
\!\!
\raisebox{-3pt}{
\begin{tikzpicture}[scale=0.4,rotate=180]
\draw[-] (0,0) -- (0.5,0.2887) -- (0.5,0.866);
\draw[-] (0.5,0.2887) -- (1,0);
\end{tikzpicture}
}
\!\!,\!\!
\raisebox{-3.8pt}{
\begin{tikzpicture}[scale=0.4,rotate=180]
\draw[-] (0,0) -- (0.5,0.2887) -- (0.5,0.866);
\draw[-] (0.5,0.2887) -- (1,0);
\draw[-,red,line width=2] (0,0) -- (0.5,0.2887) -- (0.5,0.866);
\end{tikzpicture}
}
\!\!,\!\!
\raisebox{-3pt}{
\begin{tikzpicture}[scale=0.4,rotate=180]
\draw[-] (0,0) -- (0.5,0.2887) -- (0.5,0.866);
\draw[-] (0.5,0.2887) -- (1,0);
\draw[-,red,line width=2] (0,0) -- (0.5,0.2887) -- (1,0);
\end{tikzpicture}
}
\!\!,\!\!
\raisebox{-3.8pt}{
\begin{tikzpicture}[scale=0.4,rotate=180]
\draw[-] (0,0) -- (0.5,0.2887) -- (0.5,0.866);
\draw[-] (0.5,0.2887) -- (1,0);
\draw[-,red,line width=2] (0.5,0.866) -- (0.5,0.2887) -- (1,0);
\end{tikzpicture}
}
\!\!
\big\}$
on the other. In this case the Ising interaction ferromagnetically glues together black-to-black and red-to-red, leading to closed loop states on the honeycomb lattice. In the antiferromagnetic case $J_\alpha>0$, we follow Ref.~\onlinecite{Verresen22} in labeling the four basis states as $\big\{
\!\!
\raisebox{-2pt}{
\begin{tikzpicture}[scale=0.4]
\draw[-] (0,0) -- (1,0) -- (0.5,0.866) -- (0,0);
\end{tikzpicture}
}
\!\!,\!\!
\raisebox{-2pt}{
\begin{tikzpicture}[scale=0.4]
\draw[-] (0,0) -- (1,0) -- (0.5,0.866) -- (0,0);
\filldraw[red,rotate around={60:(0.5/2,0.866/2)}] (0.5/2,0.866/2) ellipse (0.45 and 0.15);
\end{tikzpicture}
}
\!\!,\!\!
\raisebox{-4pt}{
\begin{tikzpicture}[scale=0.4]
\draw[-] (0,0) -- (1,0) -- (0.5,0.866) -- (0,0);
\filldraw[red] (0.5,0) ellipse (0.45 and 0.15);
\end{tikzpicture}
}
\!\!,\!\!
\raisebox{-2pt}{
\begin{tikzpicture}[scale=0.4]
\draw[-] (0,0) -- (1,0) -- (0.5,0.866) -- (0,0);
\filldraw[red,rotate around={-60:(0.75,0.866/2)}] (0.75,0.866/2) ellipse (0.45 and 0.15);
\end{tikzpicture}
}
\!\!
\big\}$ on one sublattice and as
$\big\{
\!\!
\raisebox{-3pt}{
\begin{tikzpicture}[scale=0.4,rotate=180]
\draw[-] (0,0) -- (1,0) -- (0.5,0.866) -- (0,0);
\end{tikzpicture}
}
\!\!,\!\!
\raisebox{-3pt}{
\begin{tikzpicture}[scale=0.4,rotate=180]
\draw[-] (0,0) -- (1,0) -- (0.5,0.866) -- (0,0);
\filldraw[red,rotate around={60:(0.5/2,0.866/2)}] (0.5/2,0.866/2) ellipse (0.45 and 0.15);
\end{tikzpicture}
}
\!\!,\!\!
\raisebox{-3pt}{
\begin{tikzpicture}[scale=0.4,rotate=180]
\draw[-] (0,0) -- (1,0) -- (0.5,0.866) -- (0,0);
\filldraw[red] (0.5,0) ellipse (0.45 and 0.15);
\end{tikzpicture}
}
\!\!,\!\!
\raisebox{-3pt}{
\begin{tikzpicture}[scale=0.4,rotate=180]
\draw[-] (0,0) -- (1,0) -- (0.5,0.866) -- (0,0);
\filldraw[red,rotate around={-60:(0.75,0.866/2)}] (0.75,0.866/2) ellipse (0.45 and 0.15);
\end{tikzpicture}
}
\!\!
\big\}$ on the other sublattice. In this case, the classical ground state degeneracy corresponds to all dimer coverings of the kagom\'e lattice. In both cases, turning on infinitesimal quantum fluctuations $\lambda$ at the isotropic point $J:=J_x=J_y=J_z$ leads to a ground state which is the equal-weight superposition of all these classical states; see Fig.~\ref{fig:kitaev}. For $J<0$ we can call this the toric code state on the honeycomb lattice \cite{Kitaev03} whereas for $J>0$ it resembles a fixed-point dimer liquid \cite{RK,Sachdev92,Sachdev99,Moessner_2001,Moessner01b,Misguich02,Fradkin_2013} on the kagom\'e lattice.

\begin{tikzpicture}

\end{tikzpicture}

Having explored the $\lambda \to +\infty$ and $\lambda \to 0$ limits, the phase diagram for $\phi=\frac{\pi}{4}$ is showcased in Fig.~\ref{fig:kitaev} for five representative values of $\lambda$. Let us first focus on the isotropic case $|J_x|=|J_y|=|J_z| = \frac{1}{3}$, which was recently studied\footnote{Ref.~\onlinecite{Verresen22} also discusses how this can be related to the Yao-Kivelson model \cite{Yao07}.} in Ref.~\onlinecite{Verresen22}: as we decrease $\lambda$, the non-Abelian phase has a transition at $\lambda_c = \frac{1}{\sqrt{3}}$, below which we enter the small-$\lambda$ phase discussed above, i.e., the toric code on the honeycomb lattice ($J_\alpha<0$) or the kagom\'e dimer liquid ($J_\alpha>0$). This raises the question of what happens for the anisotropic model: is this $\mathbb Z_2$ spin liquid connected to one of the $A_\alpha$ phase(s) of the Kitaev model? No: translation symmetry acts differently on the anyons---the $A_\alpha$ phases exhibit `weak' translation symmetry breaking since the $e$- and $m$-anyons live on alternating rows of the honeycomb lattice \cite{Kitaev06}, whereas in the low-$\lambda$ phase the hexagons only support $m$-anyons. We say they form distinct `symmetry-enriched topological' (SET) phases. Remarkably, Fig.~\ref{fig:kitaev} shows that any interpolation between these two $\mathbb Z_2$ SETs gives rise to an intermediate non-Abelian chiral phase.

In Sec.~\ref{subsec:BH} we explore an alternative (generalized) Ising model for realizing the Kitaev model. This will no longer be exactly solvable, but its interactions are arguably more straightforward to experimentally realize.

\section{Experimental relevance and generalizations \label{sec:exp}}

\subsection{General comments}
The fact that any quantum spin model can be obtained from the large-field limit of a (generalized) Ising model gives a new handle on quantum simulators and materials. We stress that there is considerable flexibility in how this idea can be applied. Firstly, while our results show that having (only) diagonal interactions is \emph{sufficient} to generate any effective off-diagonal interaction (at the expensive of reducing the on-site Hilbert space dimension), this does not imply that it would be problematic if a given experimental set-up \emph{also} has off-diagonal interactions. In fact, this can help: projecting these additional terms into the low-energy subspace can increase the number of available interactions, which can lessen the requirements on the original Hilbert space dimension (we will see an example of this in Sec.~\ref{subsec:rydberg}).

Secondly, thus far we explored two particular choices of fields (Eq.~\eqref{eq:4stateX} and Eq.~\eqref{eq:3statematrices}). Having such concrete choices allowed us to present general plug-and-chug formulas in Sec.~\ref{sec:general}. Moreover, these particular fields gave useful symmetry properties in Secs.~\ref{subsec:xy} and \ref{subsec:heisenberg}, and led to the solvable model in Sec.~\ref{subsec:kitaev}. However, in an experimental context one might want to explore a broader choice of fields. The key property to retain is that the field must have a doublet low-energy subspace. This means that by tuning a (strong) single-site field, one can obtain many different effective quantum spin models from the underlying interactions of a single microscopic model (we will see examples of this in Secs.~\ref{subsec:rydberg} and \ref{subsec:BH}).

We will now discuss two examples of experimental proposals. These happen to be in the context of analog quantum simulators. (In a sense that brings us full circle, since the present work is a generalization of the recently established connection between dimer models and Kitaev physics \cite{Verresen22} which in turn was inspired by recent Rydberg atom array theory \cite{Verresen21} and experiment \cite{Semeghini21}.) However, these ideas can equally well be explored in quantum materials. A particularly exciting direction for future work is that of novel Van der Waals heterostructures \cite{Geim2013,Andrei2020,Balents2020}, which enjoy a great degree of tunability and where (effective) higher-state models can naturally arise \cite{Po18,Wu19,Bultinck20,Zhang21}.

\subsection{Rydberg atom tweezer arrays \label{subsec:rydberg}}

\begin{figure}
\centering
\begin{tikzpicture}
\path[inner color=red!30,outer color=white] (0.75,0) ellipse (0.6 and 0.7);
\node[red] at (0.75,0.8) {$U_{nn}$};
\node at (0,0){
	\begin{tikzpicture}
	\draw[-,line width=2] (0,0) -- (1,0) node[midway,above,yshift=-2] {g.s.};
	\draw[-,black!40] (0,1.15) -- (1,1.15) node[midway,below,yshift=2] {$\scriptstyle (n-1)P$};
	\draw[-,line width=2] (0,1.7) -- (1,1.7) node[midway,above,yshift=-2] {$nS$};
	\draw[<->,black,opacity=0.2] (0.9,1.8) -- (0.9,2.15);
	\draw[-,black!40] (0,2.2) -- (1,2.2) node[midway,above,yshift=-2] {$\scriptstyle nP$};
	\draw[-,line width=2,white] (0,3) -- (1,3) node[midway,above,yshift=-2] {$\tilde nS$};
	\end{tikzpicture}
};
\node at (1.5,0){
	\begin{tikzpicture}
	\draw[-,line width=2] (0,0) -- (1,0) node[midway,above,yshift=-2] {g.s.};
	\draw[-,black!40] (0,1.15) -- (1,1.15) node[midway,below,yshift=2] {$\scriptstyle (n-1)P$};
	\draw[<->,black,opacity=0.2] (0.1,1.2) -- (0.1,1.6);
	\draw[-,line width=2] (0,1.7) -- (1,1.7) node[midway,above,yshift=-2] {$nS$};
	\draw[-,black!40] (0,2.2) -- (1,2.2) node[midway,above,yshift=-2] {$\scriptstyle nP$};
	\draw[-,line width=2,white] (0,3) -- (1,3) node[midway,above,yshift=-2] {$\tilde nS$};
	\end{tikzpicture}
};
\path[inner color=red!30,outer color=white] (4.25,0.7) ellipse (0.7 and 0.9);
\node[red] at (4.25,-0.25) {$U_{n\tilde n}$};
\node at (3.5,0){
	\begin{tikzpicture}
	\draw[-,line width=2] (0,0) -- (1,0) node[midway,above,yshift=-2] {g.s.};
	\draw[-,black!40] (0,1.15) -- (1,1.15) node[midway,below,yshift=2] {$\scriptstyle (n-1)P$};
	\draw[-,line width=2] (0,1.7) -- (1,1.7) node[midway,above,yshift=-2] {$nS$};
	\draw[<->,black,opacity=0.2] (0.9,1.8) -- (0.9,2.15);
	\draw[-,black!40] (0,2.2) -- (1,2.2) node[midway,above,yshift=-2] {$\scriptstyle nP$};
	\draw[-,black!40] (0,2.6) -- (1,2.6) node[midway,above,yshift=-2] {$\scriptstyle (\tilde n-1)P$};
	\draw[-,line width=2] (0,3) -- (1,3) node[midway,above,yshift=-2] {$\tilde nS$};
	\end{tikzpicture}
};
\node at (5,0){
	\begin{tikzpicture}
	\draw[-,line width=2] (0,0) -- (1,0) node[midway,above,yshift=-2] {g.s.};
	\draw[-,black!40] (0,1.15) -- (1,1.15) node[midway,below,yshift=2] {$\scriptstyle (n-1)P$};
	\draw[-,line width=2] (0,1.7) -- (1,1.7) node[midway,above,yshift=-2] {$nS$};
	\draw[-,black!40] (0,2.2) -- (1,2.2) node[midway,above,yshift=-2] {$\scriptstyle nP$};
	\draw[-,black!40] (0,2.6) -- (1,2.6) node[midway,above,yshift=-2,xshift=3] {$\scriptstyle (\tilde n-1)P$};
	\draw[<->,black,opacity=0.2] (0.05,2.62) -- (0.05,2.95);
	\draw[-,line width=2] (0,3) -- (1,3) node[midway,above,yshift=-2] {$\tilde nS$};
	\end{tikzpicture}
};
\node at (7,0){
	\begin{tikzpicture}
	\draw[-,line width=2] (0,0) -- (1,0) node[midway,above,yshift=-2] {g.s.};
	\draw[-,line width=2] (0,1.7) -- (1,1.7) node[midway,above,yshift=-2] {$nS$};
	\draw[-,line width=2] (0,3) -- (1,3) node[midway,above,yshift=-2] {$\tilde nS$};
	\draw[<->,color=green!80!black,line width=1.5] (0.1,0.1) -- (0.1,1.6) node[midway,right] {$\lambda$};
	\draw[<->,color=green!80!black,line width=1.5] (0.1,1.8) -- (0.1,2.9) node[midway,right] {$\lambda$};
	\draw[<->,color=green!80!black,line width=1.5] (0.9,0.1) -- (0.9,2.9) node[midway,left,yshift=-3] {$\lambda$};
	\end{tikzpicture}
};
\node at (-0.3,1.5) {(a)};
\node at (2.6,1.5) {(b)};
\node at (6.2,1.5) {(c)};
\end{tikzpicture}
\caption{\textbf{Generalized Ising model in a Rydberg tweezer array.} (a) In experiments on Rydberg atom tweezer arrays \cite{Endres16,Bernien17,Browaeys_2020,Kaufman2021}, one often encodes a qubit in the ground state (g.s.) and a Rydberg state $nS$. Virtual photons couple to the large transition dipole moment of the latter, inducing an Ising (i.e., diagonal) interaction $\sim U_{nn}$. (b) We propose to realize a generalized (here three-state) Ising model by including an extra Rydberg state $\tilde n S$. In addition to $U_{nn}$ and $U_{\tilde n \tilde n}$ (as in (a)), we also obtain a diagonal interaction $U_{n\tilde n}$. If $|n-\tilde n|>1$, there is no off-diagonal interaction. (c) Driving a strong single-site field can (effectively) couple the three levels. As derived in Sec.~\ref{subsec:3}, large $\lambda$ will transmute the diagonal Ising interactions into a spin-$1/2$ quantum magnet \eqref{eq:Rydbergspinhalf}.}
\label{fig:rydberg}
\end{figure}
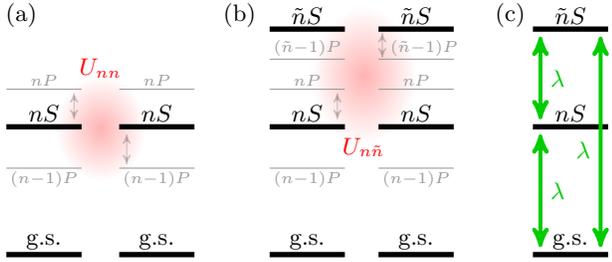

In the introduction, we already mentioned how Rydberg atom tweezer arrays \cite{Endres16,Bernien17,Browaeys_2020,Kaufman2021} can naturally realize a quantum Ising model. More precisely, the effective spin-$1/2$ is defined by the ground state $\ket{\textrm{g.s.}}$ and a highly excited Rydberg state $\ket{nS}$ of a trapped Alkali atom. If two nearby atoms are in the $\ket{nS} \otimes \ket{nS}$ state, they experience an interaction energy\footnote{There is a spatial dependence $\sim 1/r^6$ which we suppress for notational convenience. Moreover, in a wide range of circumstances it is sufficient to focus on nearest-neighbor interactions; however, sometimes longer-range terms can be important \cite{Samajdar20,Verresen21,Semeghini21,Samajdar_2021,ORourke22}.} $U_{nn} \sim n^{11}$ (see Fig.~\ref{fig:rydberg}(a)). Since lasers give us an arbitrarily tunable single-site field, we arrive at the Ising model in a transverse and longitudinal field.

A minimal change for obtaining a generalized Ising model is to consider a second Rydberg level $\ket{\tilde n S}$. By virtue of the above discussion, if two nearby atoms are in the $\ket{\tilde nS} \otimes \ket{\tilde nS}$ state, they experience an energy cost $U_{\tilde n \tilde n}$. Moreover, the state $\ket{nS} \otimes \ket{\tilde nS}$ (and its mirror) generically also gives a nonzero energy $U_{n \tilde n}$ (see Fig.~\ref{fig:rydberg}(b)). If $|n-\tilde n|>1$, there are no off-diagonal terms\footnote{This is due to $nP \neq (\tilde n-1)P$ in Fig.~\ref{fig:rydberg}(b) where $n < \tilde n$. Off-diagonal interactions must thus couple through higher levels, making them typically negligible.}.

In conclusion, we have a three-state Ising model for the qutrit $\{\ket{\textrm{g.s.}}, \ket{nS}, \ket{\tilde nS} \}$. It is a straightforward exercise (see Appendix~\ref{app:rydberg}) to write the two-body terms in terms of the diagonal qutrit operator $\mathcal Z$ defined in Eq.~\eqref{eq:3statematrices} (with $\phi \neq 0$):
\begin{equation}
H_\textrm{int} = \sum_{\langle i ,j\rangle} \left( J_{+-} \mathcal Z_i^\dagger \mathcal Z_j + J_{++} \mathcal Z_i \mathcal Z_j \right) + h.c. \label{eq:ZZfromRydberg}
\end{equation}
where
\begin{align}
9J_{+-} &= U_{nn} + U_{\tilde n \tilde n} - U_{n\tilde n} \; \; \in \mathbb R, \label{eq:Jpm}\\
9J_{++} &= \big| e^{\frac{2\pi i}{3}} U_{nn} + e^{-\frac{2\pi i}{3}} U_{\tilde n \tilde n} + 2 U_{n \tilde n} \big| \; \; \geq 0. \label{eq:Jpp}
\end{align}
By virtue of the general results in Sec.~\ref{subsec:3}, we can transmute these diagonal qutrit interactions into a spin-$1/2$ quantum magnet by introducing the (laser-induced) field $\lambda \sum_i \left(\mathcal X + \mathcal X^\dagger \right)$ defined in Eq.~\eqref{eq:3statematrices}. In the limit of large $\lambda$, Eq.~\eqref{eq:3thm} tells us we can replace $\mathcal Z_i \to \sigma^+_i$, giving the effective spin-$1/2$ model (up to single-site field which can be tuned):
\begin{equation}
H = \sum_{\langle i,j\rangle} \left( J_{+-} \sigma^+_i \sigma^-_j + J_{++} \sigma^+_i \sigma^+_j \right) + h.c. \label{eq:Rydbergspinhalf}
\end{equation}
with the coupling constants given in Eqs.~\eqref{eq:Jpm} and \eqref{eq:Jpp}.

The values of $U_{nn}, U_{\tilde n \tilde n}, U_{n \tilde n}$ (and thus $J_{+-},J_{++}$) depend on atomic physics, such as choices of $n$ and the Alkali atom. For generic choices, both $J_{++}$ and $J_{+-}$ will be nonzero and comparable. However, using the Python package \textsc{arc} \cite{arc}, we calculate (see Appendix~\ref{app:rydberg}) that choosing, e.g., Potassium atoms \cite{Angonga21} with $n=56$ and $\tilde n = 58$ gives $|J_{+-}| < 0.003 J_{++}$ (similarly for $n=89$ and $\tilde n=92$). Hence, we obtain a pure pair-creation Hamiltonian $H \propto \sum_{\langle i,j \rangle} \sigma^+_i \sigma^+_j + h.c.$. In fact, for a bipartite lattice (e.g., square or honeycomb lattice), this is unitarily equivalent to the spin-$1/2$ XY model, whereas for non-bipartite lattices (e.g., triangular or kagome lattice) this is a distinct strongly-interacting model. We provide more details (including for Rydberg and Cesium atoms) in Appendix~\ref{app:rydberg}.

The above is primarily an illustrative example of how the general approach in Sec.~\ref{sec:general} can be used to implement quantum magnets in Rydberg atom tweezer arrays. Nevertheless, even this minimal case of implementing an XY-type model has certain advantages over alternative methods. In particular, there is a natural spin-flop Hamiltonian if one encodes a spin-$1/2$ in a $\{ \ket{nS} , \ket{nP} \}$ qubit \cite{Browaeys_2020}, but this leads to $\sim 1/r^3$ dipolar tails which can be challenging for exotic states of matter with small energy gaps, and in addition, the spatial anisotropy of $p$-states prevent a direct 3D implementation. In contrast, our effective quantum magnet has much more rapidly decaying $\sim 1/r^6$ Van der Waals corrections, and it can be used for 3D geometries\footnote{These two advantages are also shared by a $\{ \ket{nS},\ket{(n+1)S} \}$ encoding \cite{Signoles21}, although this comes with $XXZ$ anisotropy and has not been shown to implement pure $XY$-interaction.}. However, by far the biggest advantage is the tunability. For instance, it is straightforward to tune single-site fields, and we can thus consider, e.g., the following field for our above Rydberg-encoded qutrit:
\begin{equation}
\lambda \left( \begin{array}{ccc}
-\sin^2 \theta & \cos^2 \theta & \cos \theta \\
\cos^2 \theta & -\sin^2 \theta & \cos \theta \\
\cos \theta & \cos \theta & 0 \end{array} \right).
\end{equation}
For $\theta=0$ this reduces to the field we used to transform the diagonal Rydberg interactions \eqref{eq:ZZfromRydberg} into the spin-$1/2$ Hamiltonian \eqref{eq:Rydbergspinhalf}. For $\theta=\frac{\pi}{2}$, large $\lambda$ instead projects out the $\ket{\tilde nS}$ state, where we thus recover the usual spin-$1/2$ Ising Hamiltonian. Hence, in the large $\lambda$ regime, one obtains an effective spin-$1/2$ model with a free parameter $\theta$ which tunes between XY and Ising interactions! It is a remarkable property of our mechanism that tuning a laser can lead to such tunable quantum spin interactions.

More generally, the results in Sec.~\ref{sec:general} can be used to realize a wide variety of quantum magnets. Firstly, if one sets $|n-\tilde n|=1$, the qutrit model will also have off-diagonal interactions; projecting these into the effective qubit space (for large $\lambda$) gives a term proportional to $\frac{1}{3} \sigma^x_i \sigma^x_j + \frac{1}{4}\sigma^z_i \sigma^z_j$, thus shifting $J_{+-}$ and $J_{++}$ equally, and introducing an XXZ anisotropy. Secondly, including yet another Rydberg level gives us a 4-state spin, and Sec.~\ref{subsec:4} shows us how its diagonal interactions can be used to realize arbitrary spin interactions. Thirdly, if one desires spatially anisotropic spin interactions (like the Kitaev honeycomb model \cite{Kitaev06}), one can leverage the anisotropic Van der Waals interactions of $p$-states. In particular, using a 4-state spin encoded in $\{ \ket{\textrm{g.s.}},\ket{nP_x},\ket{n P_y},\ket{n P_z} \}$ can be used to simulate spin-$1/2$ quantum magnets arising from strong spin-orbit coupling \cite{Jackeli09}. It would be interesting to characterize and explorate these effective Hamiltonians in future work.

\subsection{Kitaev model from the Bose-Hubbard model \label{subsec:BH}}

Finally, we show how the general results of Sec.~\ref{sec:general} can be used in the context of physical systems which might not obviously look like Ising models. In addition, we will illustrate how one can leverage the freedom in choosing the on-site field to make the results of this work broadly applicable.

In this section, we consider the case of an effective 4-state spin. In Sec.~\ref{subsec:4}, we saw how the complex-valued field \eqref{eq:4stateX} can transform diagonal interactions into off-diagonal ones. In particular, it maps the diagonal operator $\mathcal Z = e^{i\phi}\textrm{diag}(1,i,-1,-i) \to \sigma^+$. However, by changing the field direction, one can change this correspondence. In Appendix~\ref{app:4stategeneralized} we show a whole continuous family of such fields. Here we would like to focus on one particular field direction identified there:
\begin{equation}
\tilde{\mathcal X} =
\left( \begin{array}{cccc}
0 & a & a & a\\
a & 2 & \bar b & b \\
a & b & 2 & \bar b \\
a & \bar b & b & 2 \end{array}\right) \textrm{ with }
\left\{ \begin{array}{ccl}
a &= & \sqrt{1+\sqrt{3}}, \\
b &= & \sqrt{2+\sqrt{3}} e^{i\frac{5\pi}{12}}.
\end{array} \right. \label{eq:Xgen}
\end{equation}
To understand why this field direction is so useful, let us label the basis states of our 4-state spin as $\{ \ket{c}, \ket{x}, \ket{y}, \ket {z} \}$. Let $n_{\alpha=x,y,z} = 0,1$ denote whether the state $\ket{\alpha}$ is occupied (i.e., $n_\alpha \ket{\beta} = \delta_{\alpha,\beta} \ket{\beta}$).
The useful property of the field \eqref{eq:Xgen} is that when it dominates, it projects $n_\alpha \to \sigma^\alpha$ (up to a constant). This means if that one starts with the following 4-state Ising model on the honeycomb lattice:
\begin{equation}
H = V \sum_{\alpha=x,y,z} \sum_{\langle i,j\rangle_\alpha} n_{\alpha,i} n_{\alpha,j} + \lambda \sum_i \tilde{\mathcal X}_i \label{eq:nntoKitaev}
\end{equation}
(where we use the $x,y,z$ labeling of bonds as in Fig.~\ref{fig:kitaev}), then in the $\lambda \to +\infty$ limit, we obtain the Kitaev honeycomb model (up to a tunable field)!

Let us show how Eq.~\eqref{eq:nntoKitaev} can arise from a more familiar Hamiltonian, like the Bose-Hubbard model:
\begin{equation}
H_\textrm{BH} = \sum_{\langle i ,j\rangle} t_{i,j} b_i^\dagger b_j + \sum_{i,j} U_{i,j} n_i n_j + \sum_i \mu_i n_i. \label{eq:BH}
\end{equation}
In addition to the usual on-site repulsive interaction $U \equiv U_{i,i}$, it will be convenient to also include (only) a nearest-neighbor interaction $V \equiv \frac{U_{i,j}}{2}$ (for nearest neighbors $i$ and $j$). We note that this is not essential: in Appendix~\ref{app:bosehub} we show how $V$ can be perturbatively generated from hopping and the on-site interaction $U$; however, in that case the coupling constants of the eventual Kitaev honeycomb model will be smaller, and it is thus experimentally advantageous to have nonzero $V$ at the outset. We will first explore the physics of this model, and then discuss potential experimental realizations.

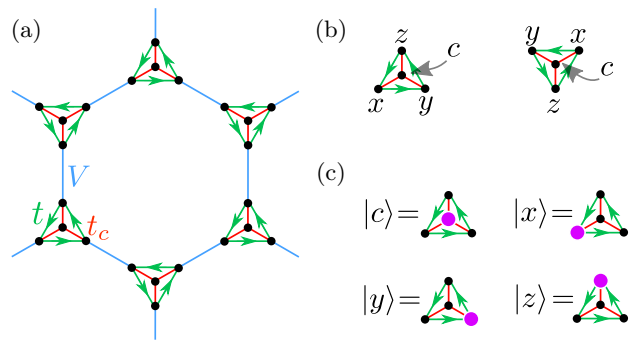
\begin{figure}
\begin{tikzpicture}
\node at (0,0) {
\includegraphics[scale=0.45]{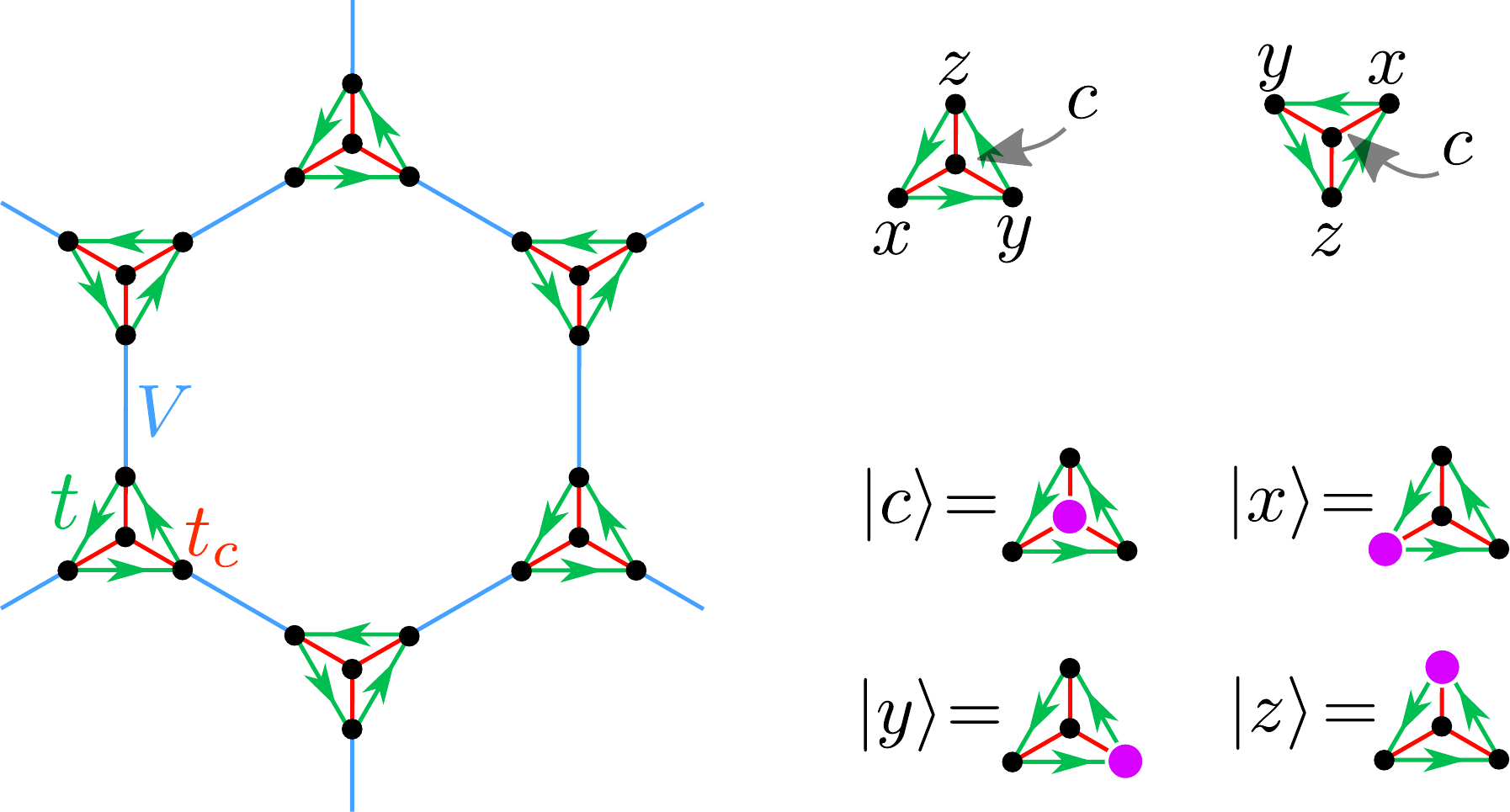}
};
\node at (-3.9,1.9) {(a)};
\node at (0.15,1.9) {(b)};
\node at (0.15,0) {(c)};
\end{tikzpicture}
\caption{\textbf{Kitaev magnet from Bose-Hubbard (BH) model.} (a) We consider the BH model \eqref{eq:BH} on a decorated honeycomb lattice. Bosons hop only \emph{within} the triangles, with complex-valued $t$ (real $t_c$) for green (red) bonds. \emph{Between} triangles, there is a density-density interaction $V$ (blue); see main text for a model with only on-site interactions. (b)  We label the four sites within a triangle as $c,x,y,z$. (c) On each triangle we put only one boson (purple dot), thereby encoding a 4-state spin. The nearest-neighbor Hubbard interaction $V$ thus defines a generalized Ising model \eqref{eq:nntoKitaev} on the honeycomb lattice. In the large-hopping limit, we obtain the spin-$1/2$ Kitaev honeycomb model with $J \propto V$ (see main text). \label{fig:BH}}
\end{figure}

Our lattice will be a decorated honeycomb lattice which can be interpreted as an overlay of the honeycomb and star (or Fisher) lattices; see Fig.~\ref{fig:BH}(a). In the minimal scenario\footnote{We note that perturbative hopping between triangles is not a problem, since we would effectively stabilize a fractional Mott insulator \cite{Motrunich02,Santos04,Buonsante05,Jurgensen14,Chen16,Barter20}; see also Appendix~\ref{app:bosehub}.}, we have nonzero hopping only within each triangle of the lattice: the red bonds have a strength $t_c = a \lambda = \sqrt{1+\sqrt{3}}\lambda $ whereas the green bonds are complex with strength $t = b \lambda =  \sqrt{2+\sqrt{3}} e^{i\frac{5\pi}{12}} \lambda$. If each triangle is occupied by exactly one boson, we obtain a 4-state spin, which we label by the basis states $\{ \ket{c}, \ket{x}, \ket{y}, \ket{z} \}$ (see Fig.~\ref{fig:BH}), with the hopping and chemical potential encoding an effective field \eqref{eq:Xgen}. Then the nearest-neighbor interaction\footnote{In Fig.~\ref{fig:BH}(a), we only show a nonzero $V$ between triangles. Since each triangle is occupied by only one boson, the interaction within a triangle is inconsequential.} $V$ indeed realizes the model in Eq.~\eqref{eq:nntoKitaev}. Details are worked out in Appendix~\ref{app:bosehub}, where we show that the large-$\lambda$ limit of $H_\textrm{BH}$ gives the Kitaev honeycomb model $H = J \sum_{\alpha=x,y,z} \sum_{\langle i , j\rangle_\alpha} \sigma^\alpha_i \sigma^\alpha_j$ with $J = V/(3+\sqrt{3})^2$ (where it is worth noting that $J$ is first-order in $V$). Naturally, if the interaction $V$ is bond-dependent, we can explore the full phase diagram of the Kitaev honeycomb model. Moreover, we can tune a single-site term to achieve the non-Abelian quantum spin liquid (see Appendix~\ref{app:bosehub}).

Let us now discuss several experimental routes towards realizing this Bose-Hubbard model, focusing on cases where the nearest-neighbor interaction is explicitly present and does not need to be generated at second-order in perturbation theory.

One natural way of obtaining such extended Bose-Hubbard interactions is dipolar physics \cite{Pupillo08,Lahaye09,Chomaz22}. Indeed, experiments on optical lattices \cite{Bloch2005,Greiner2008} have observed non-onsite interactions between, e.g., atoms with magnetic dipole moments \cite{Baier16} and molecules with electric dipole moments \cite{Yan13}. Moreover, several experimental tools for generating complex-valued hopping are known \cite{Williams10,Aidelsburger11,Struck12,Jimenez12}. More recently, tweezer arrays for molecules have been developed \cite{Liu18,Anderegg19} which make it easier to create exotic lattices; this has been demonstrated even for polar molecules \cite{Zhang22,Holland22}. We note that tweezer arrays can accommodate tunnel-coupled hopping \cite{Kaufman14,Murmann15,Spar22,Young22,Yan22}, such that the ingredients necessary for realizing the extended Hubbard model on the star-honeycomb lattice in Fig.~\ref{fig:BH}(a) have been established. It would be interesting for future work to study the effect of the longer-range dipolar $\sim 1/r^3$ interactions beyond nearest-neighbors.

Alternatively, instead of using polar atoms or molecules, one can achieve the desired properties of intra-triangle hopping and inter-triangle density-density interactions in Rydberg atom tweezer arrays. For instance, suppose that on upward-(downward-)pointing triangles of the star-honeycomb lattice in Fig.~\ref{fig:BH}(a), each black dot denotes a hardcore boson encoded in a $\{ \ket{nS},\ket{nP} \}$ ($\{ \ket{\tilde nS} , \ket{\tilde nP} \}$) qubit. For generic choices of $n \neq \tilde n$, there will be no resonant flip-flop processes. We thus have (diagonal) Van der Waals interactions between triangles (similar to Sec.~\ref{subsec:rydberg}). One advantage of this scenario is that longer-range tails decay as $\sim 1/r^6$. It remains to be seen whether the necessary complex phase factors in the (intra-triangle) hopping amplitudes can be straightforwardly achieved.

\section{Outlook \label{sec:outlook}}

In this work we have considered `generalized' Ising models---lattices of $q$-state spins (i.e., qudits) which are coupled only by diagonal interactions and are subjected to a quantum-mechanical single-site field. While the usual case of 2-state spins (i.e., qubits) has restricted phenomenology due to it being stoquastic, we have seen that Ising models for 4-state spins contain \emph{all many-body qubit Hamiltonians} by taking a large-field limit. In addition, since any qudit can be embedded into multiple qubits, a key result of our work is that any Hamiltonian defined on a collection of qudits arises from a generalized Ising model. We have illustrated this general result for a variety of paradigmatic quantum magnets, where we found rich ground state phase diagrams even for small fields, opening up such generalized Ising models as a rich field of study. Moreover, we took the first steps towards proposing novel experiments by utilizing this universal property of the Ising model. For instance, this led us to a way of realizing the Kitaev honeycomb model using cold atoms or molecules in a way that is radically distinct from previous proposals \cite{Duan03,Micheli2006,Micheli07,Schmied11,Gorshkov13,Kalinowski22b,Sun22} (as evidenced, e.g., by the $C_3$ rotation symmetry of the model in Sec.~\ref{subsec:BH}).

One interesting direction for future work is to explore and characterize the whole space of field directions which can be used to transform a given set of independent diagonal matrices into a desired Pauli algebra. In Appendix~\ref{app:4stategeneralized}, we already explored such a family, but a systematic approach would be welcome. Moreover, while the present work provides a way of obtaining arbitrary \emph{qudit} Hamiltonians from Ising models, this requires embedding qudits into multiple qubits. Future work could provide efficient user-friendly correspondences where $q$-state qudit Hamiltonians arise from $q^2$-state Ising models. Indeed, there are $q^2-1$ non-trivial independent diagonal $q^2\times q^2$ matrices, which in the large-field limit can be made to map to the $q^2-1$ generators of the Lie algebra of $SU(q)$.

In addition to further developing the theoretical framework, it is also worthwhile to explore the physics of these novel generalized Ising models. In the present work, we have already seen several surprising results, such as how a 4-state Potts chain can lead to SPT phases and their exotic criticality; it would be interesting to explore the physics of these novel Potts models in two spatial dimensions. Moreover, the solvable Ising model in Sec.~\ref{subsec:kitaev} generically gave an intervening non-Abelian spin liquid when tuning between distinct symmetry-enriched $\mathbb Z_2$ spin liquids; it is unclear whether this also holds for non-integrable models. More generally, it would be interesting to study the weak-field physics of the generalized Ising models whose large-field limit reproduces known models of interest. Note that a given quantum model can correspond to multiple Ising models. E.g., in Sec.~\ref{subsec:kitaev} we noted that choosing $\phi =0$ should lead to an alternative phase diagram where the gapless Majorana cone is stable. Moreover, if one starts with quantum models with multi-body interactions (such as the toric code \cite{Kitaev03} or cluster chain \cite{Briegel01}), the corresponding Ising model will also have multi-body interactions, which can lead to interesting physics \cite{Wegner71,Hinterman72,Griffiths73,BaxterWu73,Savvidy94,Savvidy96,Xu04,Bombin08,Yoshida14,Mueller15,Vijay16}.

Finally, there is significant experimental promise which deserves further study. In Sec.~\ref{sec:exp}, we discussed how generalized Ising models can be implemented in AMO systems. For instance, Rydberg atom tweezer arrays can encode a qudit into multiple states per atom. We identified certain set-ups which are achievable in near-term experiments (most notably Sec.~\ref{subsec:rydberg}). More broadly, it would be exciting if the same ideas can be applied to quantum materials. While the Ising model has a time-honored connection to solid-state systems \cite{DeGennes63,Wang72}, a particularly promising direction is offered by Van der Waals heterostructures \cite{Geim2013,Andrei2020,Balents2020} which admit a high degree of control, and where effective Ising models \cite{Montblanch21} and higher-state descriptions \cite{Bultinck20,Zhang21} are known to arise.

\begin{acknowledgements}
The author thanks Ashvin Vishwanath for advice and encouragement at an early stage of this project, and for collaboration on a related work \cite{Verresen22}. The author also thanks Marcus Bintz, Ruihua Fan, Francisco Machado, Daniel Parker, Rahul Sahay, Pablo Sala, Norman Yao and Michael Zaletel for stimulating conversations. DMRG simulations were performed using the TeNPy Library \cite{Hauschild18}, which was inspired by a previous library \cite{Kjaell13}. The phase diagrams in Fig.~\ref{fig:kitaev} were plotted using the python-ternary package \cite{pythonternary}. The interaction strengths for the Rydberg atom proposal were calculated using the \textsc{arc} library \cite{arc}. The author is supported by the Harvard Quantum Initiative Postdoctoral Fellowship in Science and Engineering and by the Simons Collaboration on Ultra-Quantum Matter, which is a grant from the Simons Foundation (651440, Ashvin Vishwanath). This work was performed in part at the Aspen Center for Physics, which is supported by National Science Foundation grant PHY-1607611.
\end{acknowledgements}

\bibliography{main.bbl}

\onecolumngrid

\appendix

\section{Emergence of spin-1/2 XY model from 3-state Potts model \label{app:3state}}

\subsection{Arbitrary dimensions}

We analyze the Potts model in Eq.~\eqref{eq:3statePotts}, which has the property that for $\lambda \to + \infty$ it reduces to the spin-1/2 XY model. In particular, for large $\lambda$, we project each qutrit into a two-state system given by
\begin{equation}
\ket{\downarrow} = \frac{1}{\sqrt{3}} \left( \begin{array}{c} 1 \\ \omega \\ \bar \omega \end{array} \right) \qquad \textrm{and} \qquad  \ket{\uparrow} = \frac{1}{\sqrt{3}} \left( \begin{array}{c} 1 \\ \bar \omega \\ \omega \end{array} \right).
\end{equation}
Projecting $\mathcal Z$ (defined in Eq.~\eqref{eq:3statematrices} where we choose $\phi=0$) into this space, we have that $P\mathcal Z P = \ket{\uparrow}\bra{\downarrow} = \sigma^+$. Hence, in leading-order perturbation theory, the 3-state Potts model \eqref{eq:3statePotts} for large field has an effective spin-1/2 Hamiltonian
\begin{equation}
H_\textrm{eff} = P \left( J \sum_{\langle i,j\rangle} \left( \mathcal Z_i \mathcal Z_j^\dagger + h.c. \right)\right) P = J \sum_{\langle i,j\rangle} \left( \sigma_i^\dagger \sigma_j^{\vphantom \dagger} + h.c. \right) = \frac{J}{2} \sum_{\langle i,j\rangle} \left( \sigma^x_i \sigma^x_j + \sigma^y_i \sigma^y_j \right).
\end{equation}

We can similarly calculate the terms that arise at second order in perturbation theory. For this, we can introduce the intermediate `high-energy' state
\begin{equation}
\ket{0} = \frac{1}{\sqrt{3}} \left( \begin{array}{c} 1 \\ 1 \\ 1 \end{array} \right),
\end{equation}
which costs an energy $E = 3\lambda \gg |J|$, relative to $\ket{\uparrow}$ or $\ket{\downarrow}$. We have that
\begin{equation}
P \mathcal Z \ket{0} \bra{0} \mathcal Z P = \ket{\downarrow} \bra{\uparrow}, \quad P \mathcal Z^\dagger \ket{0} \bra{0} \mathcal Z^\dagger P = \ket{\uparrow} \bra{\downarrow}, \quad
P \mathcal Z \ket{0} \bra{0} \mathcal Z^\dagger P = \ket{\downarrow} \bra{\downarrow} \quad \textrm{and} \quad
P \mathcal Z^\dagger \ket{0} \bra{0} \mathcal Z P = \ket{\uparrow} \bra{\uparrow}.
\end{equation}
Passing through this intermediate `virtual' state has three consequences. Firstly, it gives rise to second-nearest-neighbor spin-flop terms:
\begin{equation}
-\frac{J^2}{3\lambda}\sum_{\langle\langle i,j \rangle\rangle} \left( \sigma^+_i \sigma^-_j + h.c. \right).
\end{equation}
Secondly, it leads to explicitly breaking $U(1)$ down to $\mathbb Z_3$ for three neighboring sites:
\begin{equation}
- \frac{J^2}{3\lambda} \sum_{\langle i,jk \rangle} \left( \sigma^+_i \sigma^+_j \sigma^+_k + h.c. \right).
\end{equation}
Lastly, it leads to additional nearest-neighbor XXZ-type interactions:
\begin{equation}
- \frac{J^2}{6\lambda} \sum_{\langle i,j \rangle} \left( \sigma^+_i \sigma^-_j + h.c. \right) - \frac{J^2}{4\lambda} \sum_{\langle i,j\rangle} \sigma^z_i \sigma^z_j.
\end{equation}

In conclusion, to second order in perturbation theory in $\lambda \gg |J|$, we have
\begin{equation}
\boxed{ H_\textrm{eff} = \frac{J_\textrm{eff}}{2} \sum_{\langle i , j\rangle} \left( \sigma^x_i \sigma^x_j + \sigma^y_i \sigma^y_j + \Delta \sigma^z_i \sigma^z_j \right) -\frac{J^2}{3\lambda}\sum_{\langle\langle i,j \rangle\rangle} \left( \sigma^+_i \sigma^-_j + h.c. \right) - \frac{J^2}{3\lambda} \sum_{\langle i,jk \rangle} \left( \sigma^+_i \sigma^+_j \sigma^+_k + h.c. \right) }\;, \label{eq:secondorder}
\end{equation}
with
\begin{equation}
J_\textrm{eff} = J - \frac{J^2}{6\lambda}
\qquad \textrm{and} \qquad
\Delta = -\frac{1}{2\lambda/J - 1/3} = -\frac{J}{2\lambda} + O((J/\lambda)^2). \label{eq:anisotropy}
\end{equation}
Observe that $J_\textrm{eff}\Delta<0$, i.e., the XXZ interactions have ferromagnetic tendencies.

\subsection{One spatial dimension}

While Eq.~\eqref{eq:secondorder} applies to arbitrary dimensions, here we discuss the resulting physics in the one-dimensional setting. If we first focus on the nearest-neighbor interactions, we observe that we have the integrable XXZ chain \cite{Baxter1985}. This is described by a Luttinger liquid with parameter $K = \frac{1}{2-\frac{2}{\pi} \arccos(\tilde \Delta)}$ \cite{Luther75}, where we have to take $\tilde \Delta = \textrm{sgn}(J_\textrm{eff}) \Delta = - |\Delta|$. A charge-3 operator (such as the last term in Eq.~\eqref{eq:secondorder}) becomes relevant when $K \geq \frac{9}{8}$. The threshold anisotropy is thus
\begin{equation}
\tilde \Delta = \cos\left( \pi - \frac{\pi}{2K} \right) = \cos\left( \pi - \frac{4\pi}{9} \right) = \cos\left( \frac{5\pi}{9} \right) \approx - 0.174.
\end{equation}
Using Eq.~\eqref{eq:anisotropy}, this corresponds to $\frac{\lambda}{|J|} \approx 3$. For $J<0$, we expect that a field of this order of magnitude should gap out the critical phase into a symmetry-breaking phase. Qualitatively this agrees with the numerical study in Ref.~\onlinecite{Dai17} although they find that the necessary field strength differs by a factor of two from our above estimate, suggesting that we would need to go to higher order in perturbation theory to quantitatively capture this transition.

However, if $J>0$, then the spin-flip operator $\sigma^+_i$ (and any odd product of these) has momentum $\pi$. Hence, the last term in Eq.~\eqref{eq:secondorder} cannot generate $\cos(3\varphi)$ in the field theory, but only a descendant thereof (e.g. $(-1)^n \cos(3\varphi) \sim \partial \cos (3\varphi)$), which cannot gap out the critical phase. Instead, the dominant perturbation allowed by translation and $\mathbb Z_3$ symmetry is $\cos(6\varphi)$, which has scaling dimension $\frac{6^2}{4K}$. This becomes relevant only for the much larger Luttinger parameter $K \geq \frac{9}{2}$. We can surmise that the Hamiltonian under consideration will never reach this value of $K$: the antiferromagnetic Potts chain has an integrable point $J=\lambda>0$, which is a Luttinger liquid with $K=\frac{3}{2}$ \cite{Saleur91,Baxter82,OBrien20}. this suggests the following picture for $J>0$: for large fields $h\to +\infty$, we have the spin-1/2 XY chain with $K=1$, and as we lower the field, $K$ slightly increases, toward the value $K=1.5$ at the integrable point $h=J$. Throughout, there is no symmetric relevant operator, suggesting a stable gapless phase. Moreover, the remainder of the phase diagram (i.e., $0<\lambda<J$) is obtained by using the Kramers-Wannier duality, implying that the gapless phase is stabilized for all $\lambda,J > 0$. This agrees with numerical observations \cite{Dai17}.

\section{4-state generalized Ising model \label{app:4state}}

\subsection{Alternative derivation of the general theorem \label{subsec:alternative}}

Here we carry out the derivation sketched in Sec.~\ref{subsec:intuition}, which also makes the connection with Sec.~\ref{subsec:4} more explicit.

Let us first introduce:
\begin{equation}
U^x = \left( \begin{array}{cccc}
0 & 0 & 0 & 1 \\
0 & 0 & i & 0 \\
0 & -i & 0 & 0 \\ 
1 & 0 & 0 & 0 \end{array} \right),
\qquad
U^y = \left( \begin{array}{cccc}
0 & 1 & 0 & 0 \\
1 & 0 & 0 & 0 \\
0 & 0 & 0 & i \\ 
0 & 0 & -i & 0 \end{array} \right),
\qquad \textrm{and} \quad
U^z =\left( \begin{array}{cccc}
0 & 0 & 1 & 0 \\
0 & 0 & 0 & -i \\
1 & 0 & 0 & 0 \\ 
0 & i & 0 & 0 \end{array} \right). \label{eq:UxUyUz}
\end{equation}
Define $\sigma^{\alpha=x,y,z} = -U^\alpha$ and $\tau^{\alpha=x,y,z} = \left(U^\alpha\right)^*$. One can straightforwardly show that these define two sets of (mutually commuting) Pauli algebras, as the notation suggests. If we define $\mathcal Z^\alpha = - \sigma^\alpha \tau^\alpha$, then we find
\begin{equation}
\mathcal Z^x = U^x \left(U^x\right)^* = \left( \begin{array}{cccc}
1 & 0 & 0 & 0 \\
0 & -1 & 0 & 0 \\
0 & 0 & -1 & 0 \\
0 & 0 & 0 & 1
\end{array} \right), \;
\mathcal Z^y = U^y \left(U^y\right)^* = \left( \begin{array}{cccc}
1 & 0 & 0 & 0 \\
0 & 1 & 0 & 0 \\
0 & 0 & -1 & 0 \\
0 & 0 & 0 & -1
\end{array} \right), \; \textrm{and} \;
\mathcal Z^z  = U^z \left(U^z\right)^*= \left( \begin{array}{cccc}
1 & 0 & 0 & 0 \\
0 & -1 & 0 & 0 \\
0 & 0 & 1 & 0 \\
0 & 0 & 0 & -1
\end{array} \right). \label{eq:Zapp}
\end{equation}
These are thus three diagonal matrices. (Moreover, they coincide with Eq.~\eqref{eq:ZxZyZz} with $\phi=\frac{\pi}{4}$.) However, if we turn on a large field
\begin{equation}
\lambda \sum_j \left( \left( U^x_j \right)^* + \left( U^y_j \right)^*  + \left( U^z_j \right)^* \right),
\end{equation}
then for $\lambda \to + \infty$ we can everywhere replace $\left(U^\alpha_j\right)^* \to - \frac{1}{\sqrt{3}}$. In particular, $\mathcal Z^\alpha = U^\alpha \left(U^\alpha\right)^* \to - \frac{U^\alpha}{\sqrt{3}} = \frac{\sigma^\alpha}{\sqrt{3}}$. We thus see that in the presence of this large field, the three diagonal matrices in Eq.~\eqref{eq:Zapp} act like effective Pauli matrices. Lastly, observe that $\mathcal X = \sum_{\alpha=x,y,z} \left(U^\alpha\right)^*$ coincides with the expression in Eq.~\eqref{eq:4stateX}. QED.

\subsection{Explicit emergence of the spin-1/2 model}

When $\lambda \to +\infty$, we thus project each 4-state spin into an effective qubit corresponding to $\mathcal X = -\sqrt{3}$. This effective qubit can be made more explicit by defining the following basis for this subspace:
\begin{equation}
\ket{\uparrow} = \frac{e^{i\phi}}{\sqrt{2}\sqrt{3+\sqrt{3}}} \left( \begin{array}{c}
- \sqrt{2+\sqrt{3}} \\
e^{i \pi/4} \\ 
\sqrt{2+\sqrt{3}} \\ 
e^{-i\pi/4}
\end{array} \right) \qquad \textrm{and} \qquad
\ket{\downarrow} = \frac{1}{\sqrt{2}\sqrt{3-\sqrt{3}}} \left( \begin{array}{c}
- \sqrt{2-\sqrt{3}} \\
e^{-i \pi/4} \\ 
- \sqrt{2-\sqrt{3}} \\ 
e^{i\pi/4}
\end{array} \right).
\end{equation}
One can confirm that the effective action of $\sqrt{3} \mathcal Z^2$ in this basis is $\sigma^z$, whereas $\sqrt{\frac{3}{2}} e^{i\phi}\mathcal Z$ projects into $\sigma^+$. Moreover, if we set $\phi = \frac{\pi}{4}$ for concreteness, we obtain the following effective actions of the $U^{x,y,z}$ operators defined in Eq.~\eqref{eq:UxUyUz}:
\begin{equation}
U^x \to - \sigma^x, \qquad U^y \to -\sigma^y, \qquad U^z \to - \sigma^z.
\end{equation}

\subsection{Symmetries of 4-state Potts model in a complex field \label{app:4statesymmetries}}

We consider the 4-state Potts model in Eq.~\eqref{eq:4stateHeisenberg}, which has a particular complex-valued field $\mathcal X$ defined in Eq.~\eqref{eq:4stateX}. This model has an anti-unitary symmetry $S_4^T = A_4 \rtimes \mathbb Z_2^T$. I.e., odd permutations are anti-unitary, whereas even permutations are unitary. The single permutations are as follows:
{\small
\begin{align}
S_{12} &= U_{12} K = \left( \begin{array}{cccc}
	0 & -i & 0 & 0 \\
	-i & 0 & 0 & 0 \\
	0 & 0 & 1 & 0 \\
	0 & 0 & 0 & -1 \end{array} \right) K, \quad
S_{13} = U_{13} K = \left( \begin{array}{cccc}
	0 & 0 & -i & 0 \\
	0 & -1 & 0 & 0 \\
	-i & 0 & 0 & 0 \\
	0 & 0 & 0 & 1 \end{array} \right) K, \quad
S_{14} = U_{14} K = \left( \begin{array}{cccc}
	0 & 0 & 0 & -i \\
	0 & 1 & 0 & 0 \\
	0 & 0 & -1 & 0 \\
	-i & 0 & 0 & 0 \end{array} \right) K \\
S_{23} &= U_{23} K = \left( \begin{array}{cccc}
	1 & 0 & 0 & 0 \\
	0 & 0 & 1 & 0 \\
	0 & 1 & 0 & 0 \\
	0 & 0 & 0 & 1 \end{array} \right) K, \quad 
S_{24} = U_{24} K = \left( \begin{array}{cccc}
	1 & 0 & 0 & 0 \\
	0 & 0 & 0 & 1 \\
	0 & 0 & 1 & 0 \\
	0 & 1 & 0 & 0 \end{array} \right) K, \quad 
S_{34} = U_{34} K = \left( \begin{array}{cccc}
	1 & 0 & 0 & 0 \\
	0 & 1 & 0 & 0 \\
	0 & 0 & 0 & 1 \\
	0 & 0 & 1 & 0 \end{array} \right) K.
\end{align}
}
Here $K$ is complex-conjugation. The above permutations are symmetries of Eq.~\eqref{eq:4stateHeisenberg}. Indeed, clearly the Potts interaction itself is invariant under any such permutation, and for the on-site field one can straightforwardly verify this by a computation. E.g.,
\begin{equation}
S_{12} \mathcal X S_{12}^\dagger = U_{12} K \mathcal X K U_{12}^\dagger = U_{12} \mathcal X^* U_{12}^\dagger
= \left( \begin{array}{cccc}
0 & -i & 0 & 0 \\
-i & 0 & 0 & 0 \\
0 & 0 & 1 & 0 \\
0 & 0 & 0 & -1 \end{array} \right) \left( \begin{array}{rrrr} 0 & 1 & 1 & 1 \\
1 & 0 & i & -i \\
1 & -i & 0 & i \\
1 & i & -i & 0
\end{array} \right)
\left( \begin{array}{cccc}
0 & i & 0 & 0 \\
i & 0 & 0 & 0 \\
0 & 0 & 1 & 0 \\
0 & 0 & 0 & -1 \end{array} \right) = \mathcal X.
\end{equation}

Even permutations (the `alternating group') are implemented in a unitary way, which can be obtained from the above expressions. E.g., cyclically permuting the first three elements corresponds to
\begin{equation}
U_{123} = S_{12} S_{23} = U_{12} U_{23}^* = \left( \begin{array}{cccc}
0 & -i & 0 & 0 \\
-i & 0 & 0 & 0 \\
0 & 0 & 1 & 0 \\
0 & 0 & 0 & -1 \end{array} \right) \left( \begin{array}{cccc}
1 & 0 & 0 & 0 \\
0 & 0 & 1 & 0 \\
0 & 1 & 0 & 0 \\
0 & 0 & 0 & 1 \end{array} \right) = 
\left( \begin{array}{cccc}
0 & 0 & -i & 0 \\
-i & 0& 0 & 0 \\
0 & 1 & 0 & 0 \\
0 & 0 & 0 & -1
\end{array} \right),
\end{equation}
and one can confirm that this indeed commutes with $\mathcal X$.

Similarly, we obtain pairwise permutations:
\begin{align}
U_{(12)(34)} &= S_{12} S_{34} = U_{12} U_{34}^*
=  \left( \begin{array}{cccc}
0 & -i & 0 & 0 \\
-i & 0 & 0 & 0 \\
0 & 0 & 1 & 0 \\
0 & 0 & 0 & -1 \end{array} \right)
\left( \begin{array}{cccc}
1 & 0 & 0 & 0 \\
0 & 1 & 0 & 0 \\
0 & 0 & 0 & 1 \\
0 & 0 & 1 & 0 \end{array} \right)
= -i \left( \begin{array}{cccc}
0 & 1 & 0 & 0 \\
1 & 0 & 0 & 0 \\
0 & 0 & 0 & i \\
0 & 0 & -i & 0
\end{array} \right), \\
U_{(13)(24)} &= S_{13} S_{24} = U_{13} U_{24}^* =
\left( \begin{array}{cccc}
0 & 0 & -i & 0 \\
0 & -1 & 0 & 0 \\
-i & 0 & 0 & 0 \\
0 & 0 & 0 & 1 \end{array} \right) 
\left( \begin{array}{cccc}
1 & 0 & 0 & 0 \\
0 & 0 & 0 & 1 \\
0 & 0 & 1 & 0 \\
0 & 1 & 0 & 0 \end{array} \right)
= -i \left( \begin{array}{cccc}
0 & 0 & 1 & 0 \\
0 & 0 & 0 & -i \\
1 & 0 & 0 & 0 \\
0 & i & 0 & 0
\end{array} \right) .
\end{align}
Note that by comparing to Eq.~\eqref{eq:UxUyUz}, we see that $U^x = i U_{(14)(23)}$, $U^y = i U_{(12)(34)}$ and $U^z = U_{(13)(24)}$.

We observe that $A_4 \subset SO(3)$ has a projective representation on a single site. To see this, note that in $A_4$, the permutation $(12)(34)$ commutes with $(13)(24)$, whereas here we find $U_{(12)(34)} U_{(13)(24)} U_{(12)(34)}^\dagger U_{(13)(24)}^\dagger = - 1$. This captures the fact that the subgroup $\mathbb Z_2 \times \mathbb Z_2 \subset A_4$ is (projectively) represented as a dihedral group. One can also observe the projective action for the cyclic permutations. E.g., in $A_4$ the product $(123)(124)$ is an order-two element, whereas here we find $U_{123} U_{124}$ to be an element of order four. In fact, $A_4$ does not have any element of this order. Indeed, this non-trivial projective representation of $A_4$ effectively defines a faithful linear representation of $SL(2,3)$ (the double cover of $A_4$).

\subsection{Exactly-solvable SPT model}

\begin{figure}
\centering
\begin{tikzpicture}
\node at (0,0) {\includegraphics[scale=0.3]{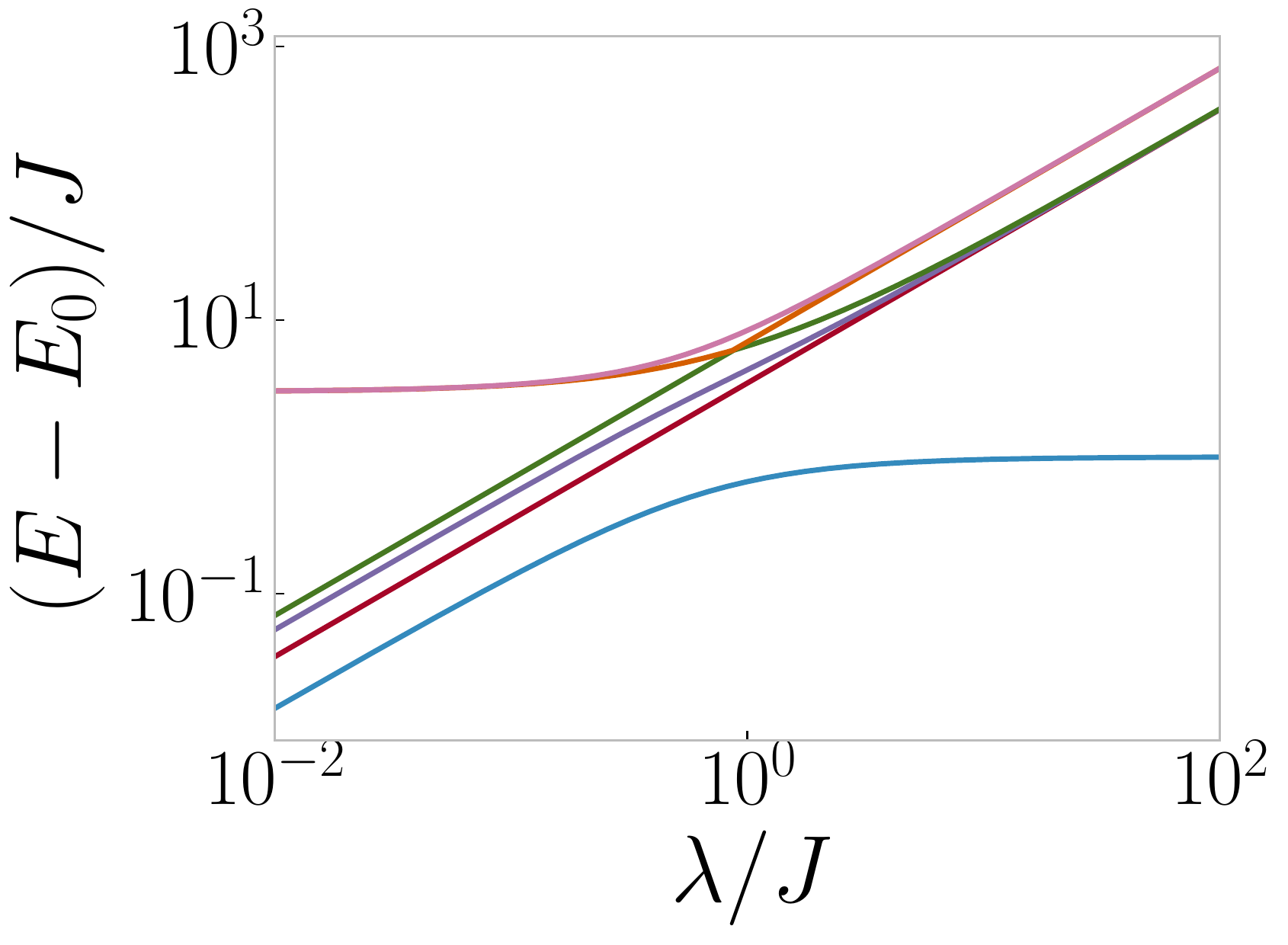}};
\node at (6,0) {\includegraphics[scale=0.3]{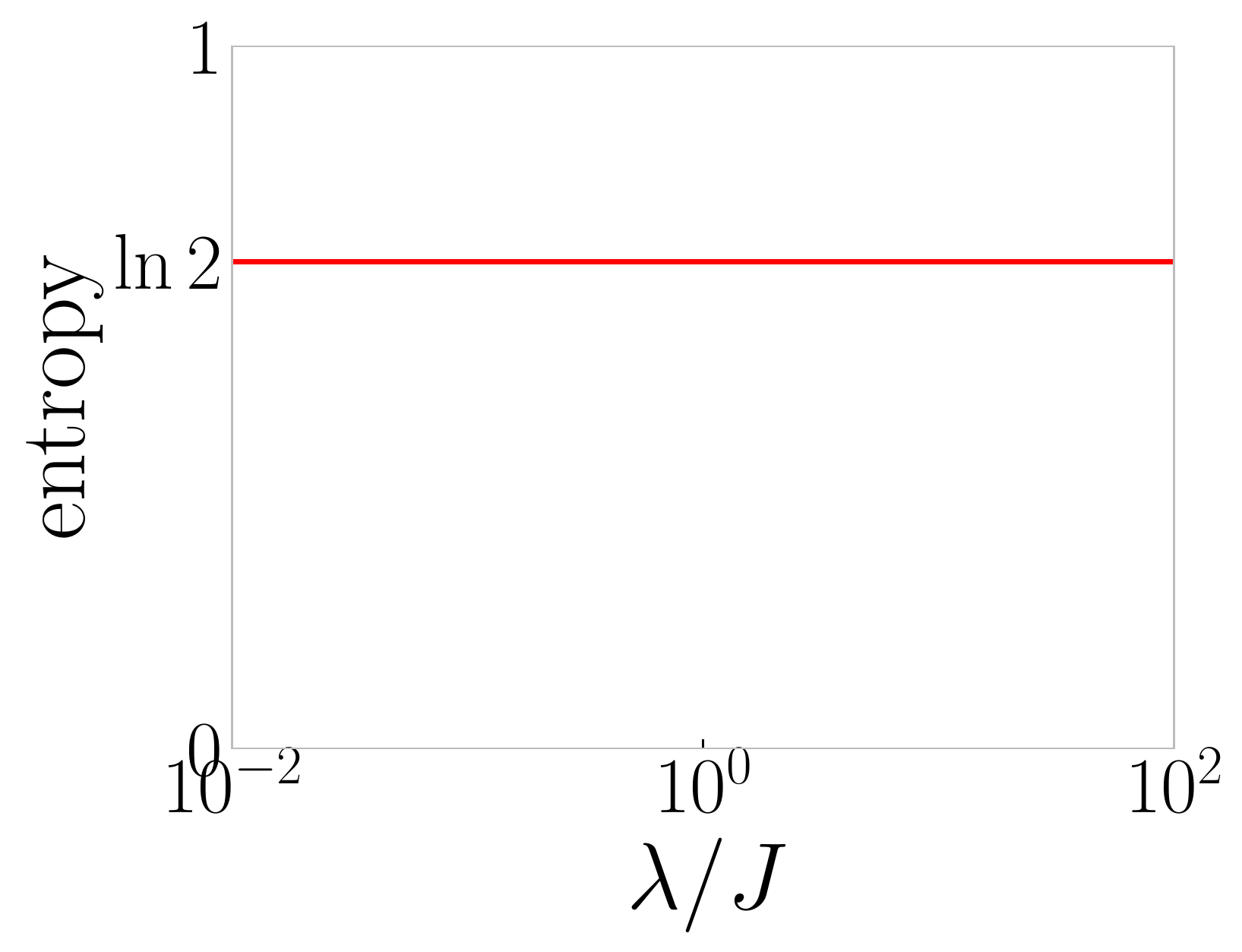}};
\node at (-2.7,1.8) {(a)};
\node at (3.5,1.8) {(b)};
\end{tikzpicture}
\caption{\textbf{Bell pair in two-site Potts model.} In the limit of strong bond-alternation of the four-state Potts model in a complex field \eqref{eq:4stateHeisenberg}, we only need to solve a two-site problem \eqref{eq:Hsolvable}. (a) We find that the ground state is gapped for all $\lambda$; this adiabatically connects to the known singlet ground state of the two-site spin-1/2 Heisenberg Hamiltonian for $\lambda \to +\infty$. (b) The ground state is a Bell pair throughout, protected by $A_4$ symmetry which is projective on a single site.}
\label{fig:SPT}
\end{figure}

In Sec.~\ref{subsec:heisenberg} of the main text, we saw that the bond-alternating 4-state Potts chain (in a complex field) effectively realizes the Haldane SPT phase. Here we briefly comment on its exactly-solvable limit, where the bond-alternation is so strong that there is no intra-unit-cell coupling (i.e., $b=1$ in the notation of the main text). In this limit, the system effectively reduces to a two-site problem:
\begin{equation}
\tilde H = 3J \delta_{1,2} + \lambda \left( \mathcal X_1 + \mathcal X_2 \right). \label{eq:Hsolvable}
\end{equation}
In the limit $\lambda \to +\infty$, this gives rise to $H = J \vec S_1 \cdot \vec S_2$. This spin-1/2 problem has four energy levels: the ground-state singlet, with a gap $J$ to three degenerate triplet states. In particular, there is a $\ln 2$ entanglement upon bipartitioning the ground state singlet. It turns out that these properties are remarkably robust. Indeed, we can solve Eq.~\eqref{eq:Hsolvable} by diagonalizing the corresponding $4^2$-dimensional matrix. The energy spectrum above the ground state is plotted in Fig.~\ref{fig:SPT}(a); we observe that there is always a nonzero energy gap. Moreover, we find that the entanglement spectrum of the ground state only contains two levels, which are exactly degenerate due to the projective $A_4$ action on a single site, as discussed in the previous subsection; this leads to the robust $\ln 2$ entanglement plotted in Fig.~\ref{fig:SPT}(b).

In conclusion, the ground state is a Bell pair for all values of $\lambda>0$ !

\section{Kitaev model from a generalized transverse-field Ising model \label{app:kitaev}}

Here we analyze
\begin{equation}
\tilde H = 3 \sum_{\alpha=x,y,z} J_\alpha \sum_{\langle i,j \rangle_\alpha} \mathcal Z^\alpha_i \mathcal Z^\alpha_j + \lambda \sum_j \mathcal X_j
\end{equation}
with
\begin{equation}
\mathcal Z^x_j = \frac{e^{i\phi} \mathcal Z_j + e^{-i \phi} \mathcal Z^\dagger_j}{\sqrt{2}}, \;
\mathcal Z^y_j = \frac{e^{i\phi} \mathcal Z_j - e^{-i \phi} \mathcal Z^\dagger_j}{\sqrt{2}i}, \;
\mathcal Z^z_j = \mathcal Z_j^2,
\end{equation}
where $\mathcal Z_j$ and $\mathcal X_j$ are defined in Eq.~\eqref{eq:4stateX}, and we take the particular choice $\phi = \frac{\pi}{4}$ for which we obtain the matrices $\mathcal Z^\alpha$ in Eq.~\eqref{eq:Zapp}.
The interest in this generalized Ising model is that in the limit $\lambda \to +\infty$, it reduces to the Kitaev honeycomb model \cite{Kitaev06}; see Eq.~\eqref{eq:general}. For ease of notation and discussion, we will set $J_\alpha>0$, but we note that this sign can be unitarily toggled.

\subsection{Free-fermion solution \label{app:freefermsol}}

Here we largely follow Ref.~\onlinecite{Verresen22}, although some of the details are worked out in a different way, and moreover, the phase diagram had so far only been obtained in the isotropic case $J_x=J_y=J_z$. Each 4-state spin can be described in terms of six Majorana operators \cite{Kitaev06,Wang09,Yao09,Yao11,Chua11,Whitsitt12,Natori16,Natori17,deCarvalho18,Natori18,Seifert20,Farias20,Natori20,Chulliparambil20,Ray21,Jin21,Chulliparambil21} with a parity condition, $ib^x b^y b^z c^x c^y c^z = 1$. The three commuting operators $\mathcal Z^\alpha$ in Eq.~\eqref{eq:Zapp} can then be expressed as $\mathcal Z^\alpha = i b^\alpha c^\alpha$; indeed note that the right-hand is hermitian and squares to the identity. Moreover, in Sec.~\ref{app:proofX} we prove that we can equate:
\begin{equation}
\mathcal X = -i\left( c^x c^y + c^y c^z + c^z c^x \right). \label{eq:Xmaj}
\end{equation}
Then $\tilde H$ can be written as
\begin{equation}
\tilde H = - 3 J_\alpha \sum_{\alpha=x,y,z} \sum_{\langle j,k\rangle_\alpha} \hat u_{jk} i c_j^\alpha c_k^\alpha - \lambda \sum_j \left(i c^x_j c^y_j +i c^y_j c^z_j + ic^z_j c^x_j \right) \qquad \textrm{with } \hat u_{jk} = i b^\alpha_j b^\alpha_k.
\end{equation}
Here $\hat u_{jk}$ is a conserved quantity. We numerically find that the ground state lies in the sector $u_{jk} = 1$, where we take the convention that it points from the A sublattice to the B sublattice. This agrees with the large-$\lambda$ limit (which reduces to the spin-1/2 Kitaev model \cite{Kitaev06} as explained in the main text) and perturbation theory in the small-$\lambda$ limit \cite{Verresen22}, and the isotropic case which is equivalent to the Yao-Kivelson model \cite{Yao07}, all of which are known to be flux-free.

We thus obtain an effective-free fermion Hamiltonian, which in momentum space is described by:
\begin{equation}
\mathcal H_{k_x,k_y} = \left( \begin{array}{ccc|ccc}
0 & i \lambda & -i \lambda & 0 & 3iJ_x e^{-ik_x} & 0 \\
-i\lambda & 0 & i \lambda & 0 & 0 & 3iJ_y e^{-ik_y} \\ 
i \lambda & - i \lambda & 0 & 3iJ_z & 0 & 0 \\ \hline
0 & 0 & -3i J_z & 0 & i\lambda &-i\lambda \\ 
-3iJ_x e^{ik_x} & 0 & 0 & -i \lambda & 0 & i\lambda \\ 
0 & -3i J_y e^{ik_y} & 0 & i\lambda & -i\lambda &0
\end{array} \right), \label{eq:Hfreeferm}
\end{equation}
where $k_x,k_y$ are coordinates in the reciprocal basis.

Let us set $\tilde J_\alpha = 3 J_\alpha$ for convenience. For $k_x, k_y \in \{0,\pi\}$ we find
\begin{equation}
\sqrt{|\textrm{det} \left( \mathcal H \right)|} = \left| e^{ik_x}\tilde J_x e^{ik_y}\tilde J_y \tilde J_z - \left(e^{ik_x}\tilde J_x + e^{ik_y}\tilde J_y + \tilde J_z \right) \lambda^2 \right|.
\end{equation}

This suggests the following picture: within the `$B$ region' of the Kitaev model (i.e., where the triangle inequalities such as $|J_x| \leq |J_y| + |J_y|$ and permutations thereof are satisfied), we have a (known) gapless phase for $\lambda \to +\infty$, which opens up into a gapped chiral spin liquid for finite $\lambda$, until we reach a critical field value
\begin{equation}
\lambda_c = \sqrt{ \frac{|\tilde J_x \tilde J_y \tilde J_z|}{|\tilde J_x| + |\tilde J_y| + |\tilde J_z|} },
\end{equation}
below which there is a gapped $\mathbb Z_2$ spin liquid, adiabatically connecting to the kagom\'e dimer liquid which is discussed in detail in Ref.~\onlinecite{Verresen22} for the isotropic case $J_x=J_y=J_z$ and small $h$.

In the $A$ region, we start with a $\mathbb Z_2$ spin liquid for large fields $\lambda \to + \infty$. For concreteness, let us focus on the $A_z$ region, where $|J_z| \geq |J_x|+|J_z|$. As we lower the field, there will be a first transition:
\begin{equation}
\lambda_{c_1} = \sqrt{ \frac{|\tilde J_x \tilde J_y \tilde J_z|}{|\tilde J_z| - |\tilde J_x| - |\tilde J_y|} },
\end{equation}
where we enter the chiral spin liquid, and as we continue to lower the field, we encounter the same transition we discussed above for the B region:
\begin{equation}
\lambda_{c_2} = \sqrt{ \frac{|\tilde J_x \tilde J_y \tilde J_z|}{|\tilde J_x| + |\tilde J_y| + |\tilde J_z|} },
\end{equation}
into the kagom\'e dimer spin liquid. Note that if we approach the $B$ region, then $\lambda_{c_1} \to + \infty$. If we instead approach one of the corners of the triangle phase diagram (i.e., $|J_z| \to +\infty$), then $\lambda_{c_1},\lambda_{c_2} \to \sqrt{| \tilde J_x \tilde J_y|}$.

By numerically diagonalizing Eq.~\eqref{eq:Hfreeferm} we have confirmed that there no transitions for $k_x,k_y \notin \{0,\pi\}$. The above analytic expressions were used to plot the representative phase diagrams in Fig.~\ref{fig:kitaev} in the main text.

\subsection{Proof of Eq.~\eqref{eq:Xmaj} \label{app:proofX}}

Let us choose an eigenbasis for $\mathcal Z^\alpha$:
\begin{equation}
ib^y c^y \ket{y,z} = \mathcal Z^y |y,z\rangle = y|y,z\rangle, \qquad i b^z c^z \ket{y,z} =  \mathcal Z^z |y,z\rangle = z |y,z\rangle, \qquad \textrm{where } y,z \in \{-1,1\}.
\end{equation}
Note that $\mathcal Z^x |y,z\rangle = yz | y,z\rangle $ since $i b^x c^x = \left( i b^y c^y \right) \left( i b^z c^z \right)$ due to the parity condition. The $c$-pairing operators toggle these basis states:
\begin{equation}
i c^x c^y \ket{y,z} = \ket{-y,z}, \label{eq:toggle}
\end{equation}
since $i c^x c^y$ anticommutes with $i b^y c^y$. Eq.~\eqref{eq:toggle} only needs to hold up to phase factors, but such phase factors can be absorbed into the (re)definition of our basis basis states. Similarly $i c^y c^z$ will toggle $\ket{y,z} \leftrightarrow \ket{-y,-z}$, however, now the phase factors are no longer completely free. To determine the phase factors, we write
\begin{equation}
i c^y c^z \ket{y,z} = g(y,z) \ket{-y,-z}.
\end{equation}

Then:
\begin{equation}
\left( i c^x c^y \right) \left( i c^y c^z \right) \ket{y,z} = g(y,z) ic^x c^y \ket{-y,-z} = g(y,z) \ket{y,-z}
\end{equation}
but also
\begin{equation}
\left( i c^x c^y \right) \left( i c^y c^z \right) \ket{y,z} = -\left( i c^y c^z \right) \left( i c^x c^y \right) \ket{y,z} = -ic^y c^z \ket{-y,z} = - g(-y,z) \ket{y,-z}.
\end{equation}
We thus conclude that $g(-y,z) = - g(y,z)$. Moreover, since the operator squares to identity, we have that $g(y,z) g(-y,-z) = 1$. This fixes the phase factors: if we set $g(1,1)=1$ then $g(-1,1) = -1 = g(1,-1)$ and $g(-1,-1) = g(1,1) = 1$.

Note that $ic^z c^x =i  \left( i c^x c^y \right) \left( i c^y c^z \right)$, hence the third $c$-pairing operator is fixed by the first two.

We can now write down the matrix representation in the basis $\{ \ket{1,1},\ket{1,-1},  \ket{-1,1},  \ket{-1,-1} \}$. Firstly,
\begin{equation}
i b^y c^y = \mathcal Z^y = \left( \begin{array}{rrrr}
1 & 0 & 0 & 0 \\
0 & 1 & 0 & 0 \\
0 & 0 & -1 & 0 \\
0 & 0 & 0 & -1
\end{array} \right) 
\qquad \textrm{and} \qquad
i b^z c^z = \mathcal Z^z = \left( \begin{array}{rrrr}
1 & 0 & 0 & 0 \\
0 & -1 & 0 & 0 \\
0 & 0 & 1 & 0 \\
0 & 0 & 0 & -1
\end{array} \right),
\end{equation}
and $\mathcal Z^x = \mathcal Z^y \mathcal Z^z$. This coincides with the matrix representation in Eq.~\eqref{eq:Zapp}. Secondly, our above analysis for the $c$-pairing operators gives us:
\begin{equation}
i c^x c^y = \left( \begin{array}{rrrr}
0 & 0 & 1 & 0 \\
0 & 0 & 0 & 1 \\
1 & 0 & 0 & 0 \\
0 & 1 &0 & 0
\end{array} \right), \quad
i c^y c^z = \left( \begin{array}{rrrr}
0 & 0 & 0 & 1 \\
0 & 0 & -1 & 0 \\
0 & -1 & 0 & 0 \\
1 & 0 &0 & 0
\end{array} \right) \quad \textrm{and} \quad
i c^z c^x = i \left( i c^x c^y\right) \left( i c^y c^z \right) =
\left( \begin{array}{rrrr}
0 & -i & 0 & 0 \\
i & 0 & 0 & 0 \\
0 & 0 & 0 & i \\
0 & 0 & -i & 0
\end{array} \right).
\end{equation}
Hence:
\begin{equation}
i c^x c^y + i c^y c^z + i c^z c^x 
= \left( \begin{array}{rrrr}
0 & -i & 1 & 1 \\
i & 0 & -1 & 1 \\
1 & -1 & 0 & i \\
1 & 1 & -i & 0
\end{array} \right).
\end{equation}

This can be equated with $\mathcal X$ in Eq.~\eqref{eq:4stateX} after a change of basis. More precisely, the matrix representation in the basis $\{ \ket{1,1}, i\ket{1,-1},-\ket{-1,1}, - \ket{-1,-1} \}$ is the same for the diagonal $\mathcal Z^\alpha$ operators, but:
\begin{equation}
i c^x c^y + i c^y c^z + i c^z c^x 
= \left( \begin{array}{rrrr}
0 & -1 & -1 & -1 \\
-1 & 0 & i & -i \\
-1 & -i & 0 & i \\
-1 & i & -i & 0
\end{array} \right) = - \mathcal X.
\end{equation}

\section{Further generalizations}

\subsection{A broader class of 4-state fields \label{app:4stategeneralized}}

In Sec.~\ref{subsec:4} of the main text we saw how a particular choice of complex field on a 4-state system can give rise to an effective spin-1/2 such that diagonal operators now act as Pauli operators. (In Appendix~\ref{app:4state} we provided more details.) Here we consider a generalization of this field:
\begin{equation}
\mathcal X(q) =  \left( \begin{array}{cccc}
-3 q & \sqrt{1-3 q^2} &  \sqrt{1-3 q^2} &  \sqrt{1-3 q^2} \\
\sqrt{1-3 q^2}& q & -i+q & i+q\\
\sqrt{1-3 q^2}& i+q & q & -i+q \\
\sqrt{1-3 q^2}& -i+q& i+q & q
\end{array} \right). \label{eq:Xgeneralized}
\end{equation}
We will consider this operator for the range $q^2 < 1/3$. Note that if we set $q=0$, then it reduces to the field in Eq.~\eqref{eq:4stateX} of the main text, i.e., $\mathcal X(0) = \mathcal X$. Also comparing to Eq.~\eqref{eq:Xgen}, we identify $\mathcal X(2-\sqrt{3}) = (2-\sqrt{3})(2\tilde{\mathcal X} - 3 \mathbb I)$.

It can be readily shown that $\mathcal X(q)$ has eigenvalues $\{-\sqrt{3}.-\sqrt{3},\sqrt{3},\sqrt{3}\}$ and thus satisfies $\mathcal X(q)^2 = 3$. If we add the above field (with a large prefactor) to a 4-state spin, we will project into the low-energy spin-1/2, using the projector $P = \frac{1}{2} \left( \mathbb I_4 - \mathcal X(q)/\sqrt{3} \right)$.

In particular, we consider the following three independent diagonal matrices:
\begin{equation}
n^x = \left( \begin{array}{cccc}
0 & 0 & 0 & 0 \\
0 & 1 & 0 & 0 \\
0 & 0 & 0 & 0 \\
0 & 0 & 0 & 0 \end{array} \right),
\qquad
n^y = \left( \begin{array}{cccc}
0 & 0 & 0 & 0 \\
0 & 0 & 0 & 0 \\
0 & 0 & 1 & 0 \\
0 & 0 & 0 & 0 \end{array} \right),
\qquad
n^z = \left( \begin{array}{cccc}
0 & 0 & 0 & 0 \\
0 & 0 & 0 & 0 \\
0 & 0 & 0 & 0 \\
0 & 0 & 0 & 1 \end{array} \right). \label{eq:n}
\end{equation}

\begin{figure}
\begin{tikzpicture}
\node at (0,0) {\includegraphics[scale=0.34]{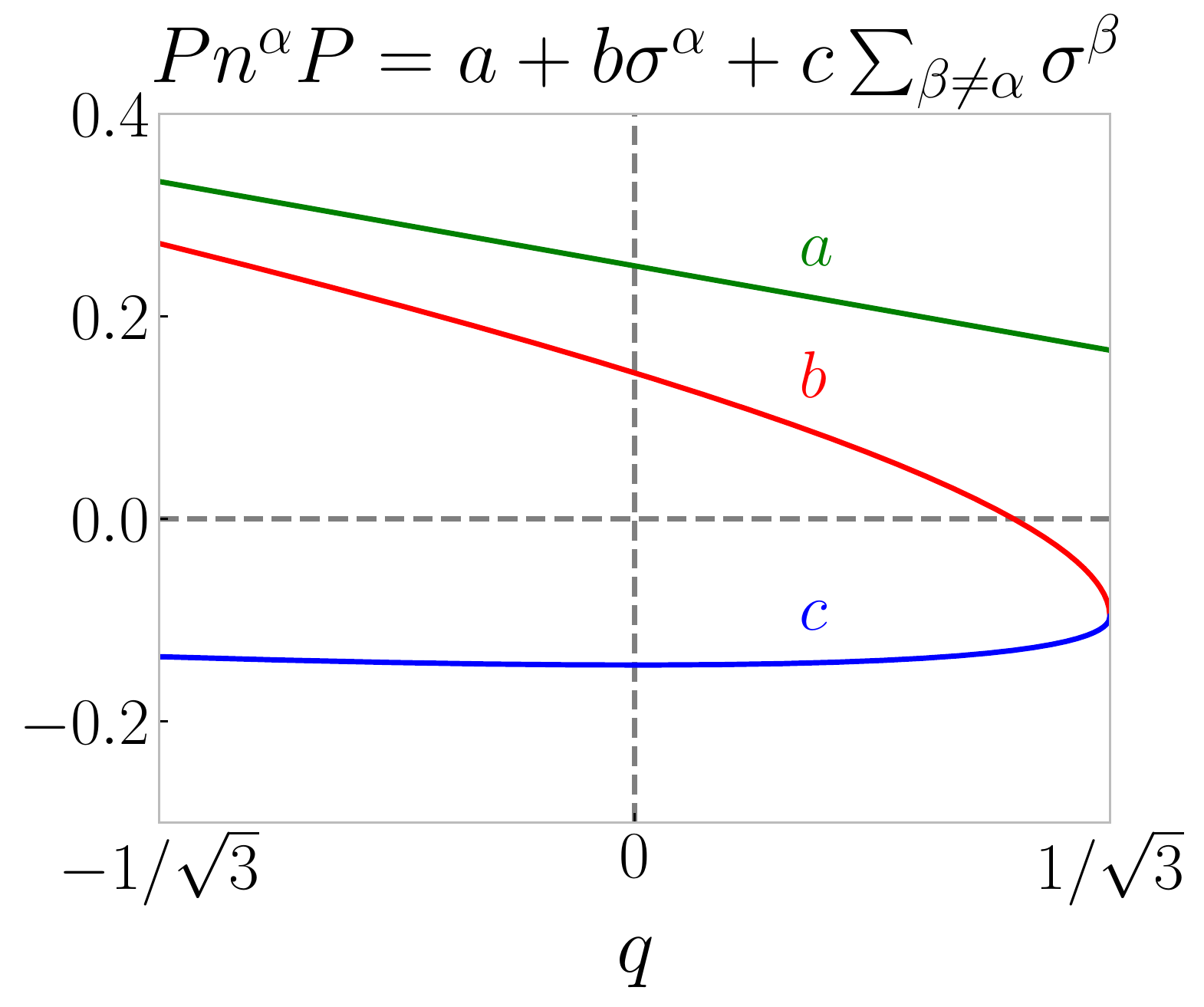}};
\node at (7,0) {\includegraphics[scale=0.34]{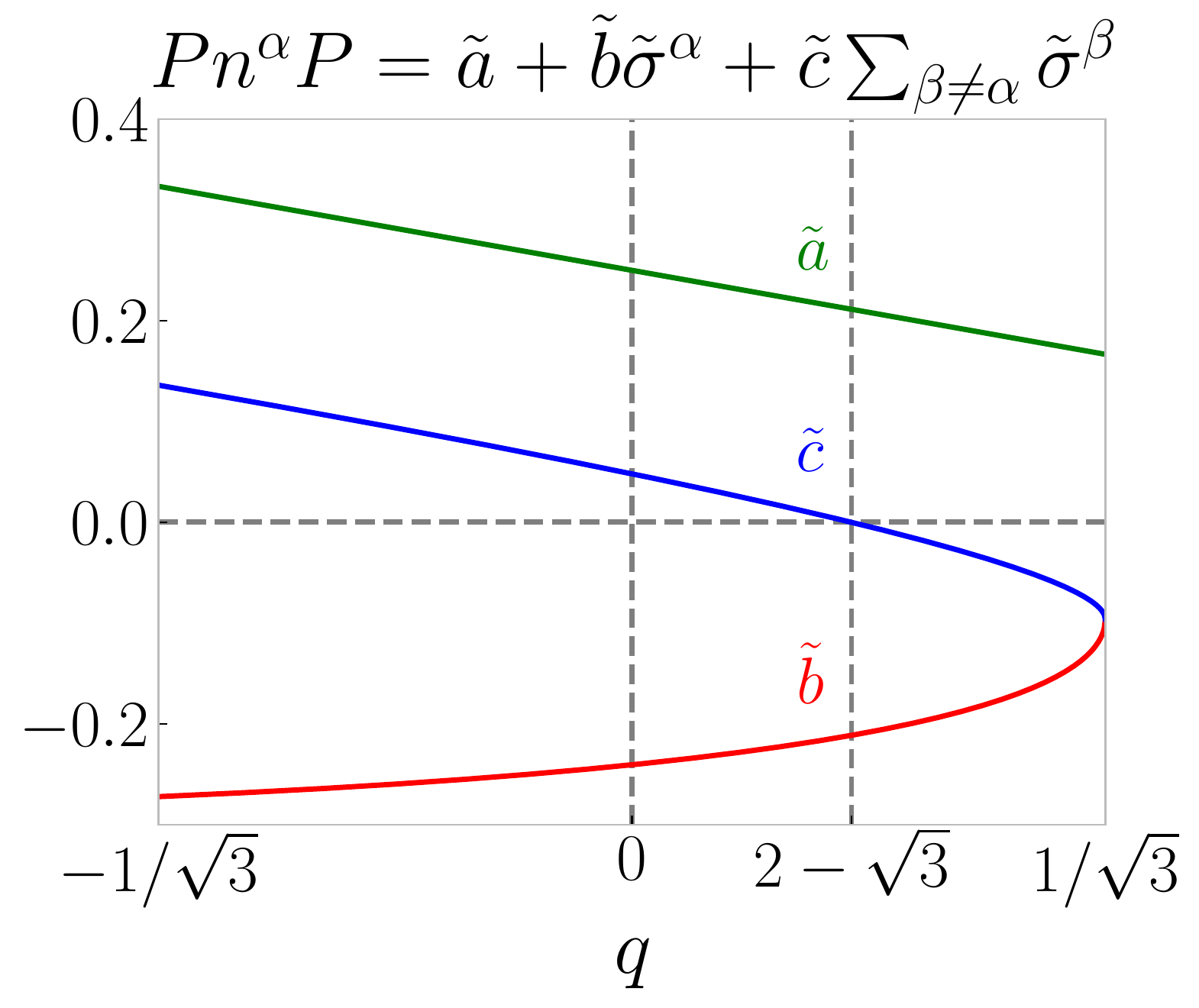}};
\node at (-3,2) {(a)};
\node at (4,2) {(b)};
\end{tikzpicture}
\caption{\textbf{Field-tuning the effective Pauli operators obtained from particle density.} We consider the 4-state field $\mathcal X(q)$ in Eq.~\eqref{eq:Xgeneralized} with free parameter $|q| < 1/\sqrt{3}$. The special case $\mathcal X(0)$ coincides with Eq.~\eqref{eq:4stateX} in the main text. We track how the three independent density operators $n^\alpha$ in Eq.~\eqref{eq:n} transform into Pauli operators in the large-field limit. (a) We show the solution which is continuously connected to the result for $q=0$ in Eq.~\eqref{eq:q0}. (b) We obtain an equivalent set of results after a change of basis (i.e., a different choice of matrices satisfying the Pauli algebra). The advantage of this basis is that at $q=2-\sqrt{3}$, only a single Pauli operator $\sigma^\alpha$ appears after projecting the diagonal operator $n^\alpha$. In Appendix~\ref{app:bosehub} we use this to encode the Kitaev interactions using nearest-neigbor density-density interactions in a Bose-Hubbard model; such interactions can be explicitly present (Sec.~\ref{subsubsec:BHinteraction}) or can perturbatively arise from hopping terms (Sec.~\ref{subsubsec:BHhopping}). \label{fig:PnP}}
\end{figure}

If we set $q=0$, we have
\begin{equation}
P n^x P = \frac{ 1 + (\sigma^x-\sigma^y-\sigma^z)/\sqrt{3}}{4}, \quad
P n^y P = \frac{ 1 + (-\sigma^x+\sigma^y-\sigma^z)/\sqrt{3}}{4}, \quad
P n^z P = \frac{ 1 + (-\sigma^x -\sigma^y+\sigma^z)/\sqrt{3}}{4}, \label{eq:q0}
\end{equation}
which is different but equivalent way of writing the properties discussed in Sec.~\ref{subsec:4} of the main text. For instance, $\mathcal Z^2 = \mathbb I_4 - 2n^x - 2n^y$, which thus projects into $1-\frac{1+(\sigma^x-\sigma^y-\sigma^z)/\sqrt{3}}{2}-\frac{1+(-\sigma^x+\sigma^y-\sigma^z)/\sqrt{3}}{2} = \sigma^z/\sqrt{3}$, as claimed in Sec.~\ref{sec:general}.

In the next subsection, we will be interested in a different choice of $q$ (which will aid in implementing the Kitaev honeycomb model using a Bose-Hubbard model): setting $q = \tan(\pi/12)= 2-\sqrt{3} \approx 0.2679$, then
\begin{equation}
P n^\alpha P = \frac{ 1 - \sigma^\alpha}{3+\sqrt{3}}. \label{eq:projn}
\end{equation}

For completeness, in Fig.~\ref{fig:PnP} we track the coefficients of the Pauli operators in $P n^\alpha P$ for the whole family of $q^2 < 1/3$. In panel (a), we start with the solution in Eq.~\eqref{eq:q0} for $q=0$, and we find the continuous family connected to it. While there is a point where $Pn^\alpha P$ only contains two of three Pauli operators (i.e., where the red curve vanishes), it does not directly seem to contain the solution shown in Eq.~\eqref{eq:projn}. It turns out that we have to perform a discrete change of basis, after which all the solutions in panel (a) are mapped to those in panel (b). In this basis, we indeed recognize that for $q=2-\sqrt{3}$, the coefficients in front of two of the three Paul operators vanish.

\subsection{Kitaev honeycomb model from Bose-Hubbard model on the star-honeycomb lattice \label{app:bosehub}}

We consider the Bose-Hubbard model on the star-honeycomb lattice shown in Fig.~\ref{fig:BH} of the main text. More precisely, each `site' of the honeycomb lattice hosts a triangle; the boson can either occupy one of the three corners (labeled $x,y,z$) or the center (labeled $c$). The Hamiltonian is of the form $H = H_\textrm{intra} + H_\textrm{inter}$, where the first piece is dominant and describes the hopping within a triangle, and the latter piece specifies how the different triangles couple. We consider two illustrative cases: in Sec.~\ref{subsubsec:BHinteraction} there is a nearest-neighbor interaction, whereas Sec.~\ref{subsubsec:BHhopping} considers the more restrictive case of on-site interactions, where the triangles are coupled only by hopping terms. In both cases, there is a limit where the effective low-energy model is the spin-1/2 Kitaev honeycomb model.

\subsubsection{Bose-Hubbard model with nearest-neighbor interaction \label{subsubsec:BHinteraction}}
We consider $H = H_\textrm{intra} + H_\textrm{inter}$ with
\begin{align}
H_\textrm{intra} &= \sum_{j} \left[ \sum_{\alpha=x,y,z} \left( t_c b_{j,\alpha}^\dagger b_{j,c}^{\vphantom \dagger} + t b_{j,\alpha+1}^\dagger b_{j,\alpha}^{\vphantom \dagger} + h.c. \right) + \sum_{a=c,x,y,z} \mu_{a} n_{j,a} \right] \\
H_\textrm{inter} &=  \sum_{\alpha=x,y,z} V_\alpha \sum_{\langle j,k\rangle_\alpha} n_{j,\alpha} n_{k,\alpha}. \label{eq:nnrepulsion}
\end{align}
By changing the separation distances between triangles, the couplings $V_x,V_y,V_z$ can in principle have distinct values. While we do not require any additional couplings, a generic Bose-Hubbard model might have more terms. Let us briefly discuss why such additional terms would not affect our discussion:
{\small
\begin{itemize}
\item There might be a nearest-neighbor interaction $\sim V$ between two sites on the \emph{same} triangle (i.e., in $H_\textrm{intra}$). However, we will only consider filling each triangle with one boson, and hence this interaction has no effect. 
\item For the same reason, an on-site Hubbard term $U n_{j,a} (n_{j,a}-1)$ has no effect.
\item There might be a boson-hopping term in $H_\textrm{inter}$. However, if there are repulsive terms $U>0$ or $V>0$ in a triangle (see previous bulletin points), then two bosons on a single triangle will have less kinetic energy lowering than if each boson had their own triangle; hence, in the limit $\lambda \to + \infty$ considered below, such boson-hopping will be projected out---at leading order in perturbation theory, only the diagonal term in $H_\textrm{inter}$ will contribute. (In the absence of the diagonal interaction in Eq.~\eqref{eq:nnrepulsion}, such hoppings can lead to the Kitaev model at second-order in perturbation theory; see Sec.~\ref{subsubsec:BHhopping}.)
\end{itemize}
}

Let us first set the inter-triangle interactions to zero (i.e., $V_x = V_y = V_z = 0$). We put a single boson on each triangle, with the following choice of parameters:
\begin{align}
t_c &= 2\lambda \sqrt{3\sqrt{3}-5} \approx 0.886 \lambda, \\
t &= \lambda(2-\sqrt{3}+i) = 2\lambda \sqrt{2-\sqrt{3}} e^{i 5\pi/12} \approx 1.035 \; e^{1.309i} \; \lambda,\\
\mu_{\alpha=x,y,z} - \mu_c &= 4 \lambda \left(2-\sqrt{3}\right) + \nu_\alpha \approx 1.072 \lambda + \nu_\alpha.
\end{align}
In the limit $\lambda \to + \infty$ (where we keep $\nu_\alpha$ finite), each triangle becomes an effective two-level system as discussed around Eq.~\eqref{eq:Xgeneralized} (here $q=2-\sqrt{3}$; note that since $\mathcal X(2-\sqrt{3}) = (2-\sqrt{3})(2\tilde{\mathcal X} - 3 \mathbb I)$ there is a proportionality factor relative to the parameters in Sec.~\ref{subsec:BH} which can be absorbed by rescaling $\lambda$). We thus obtain a spin-1/2 model on the honeycomb lattice.

Subsequently turning on $H_\textrm{inter}$ and projecting into the low-energy subspace (using Eq.~\eqref{eq:projn}), we obtain the following effective spin-1/2 honeycomb Hamiltonian:
\begin{align}
H_\textrm{eff} &= \sum_{\alpha=x,y,z} V_\alpha \sum_{\langle j,k\rangle_\alpha} \frac{ 1 - \sigma_j^\alpha}{3+\sqrt{3}} \times \frac{ 1 - \sigma_k^\alpha}{3+\sqrt{3}} + \sum_{j} \sum_{\alpha=x,y,z} \nu_\alpha \frac{ 1 - \sigma_j^\alpha}{3+\sqrt{3}} \\
&= \sum_{\alpha=x,y,z} \left( J_\alpha \sum_{\langle j,k\rangle_\alpha} \sigma^\alpha_j \sigma^\alpha_k - h_\alpha \sum_j \sigma^\alpha_j \right) + \textrm{const.} \label{eq:effectiveKitaev}
\end{align}
Here we used that $n_c+n_x+n_y+n_z=1$ (for any triangle). The effective couplings of the spin-1/2 honeycomb model are:
\begin{equation}
J_\alpha = \frac{V_\alpha}{\left(3+\sqrt{3}\right)^2} = \frac{(2-\sqrt{3})V_\alpha}{6} \approx 0.0447 \; V_\alpha \quad \textrm{and} \quad h_\alpha =
\frac{V_\alpha}{\left(3+\sqrt{3}\right)^2}  + \frac{\nu_\alpha}{3+\sqrt{3}} \approx 0.0447 V_\alpha + 0.211 \nu_\alpha.
\end{equation}
Hence, if we tune the chemical potential such that $\nu_\alpha = - \frac{V_\alpha}{3+\sqrt{3}} \approx  -0.211 V_\alpha$, we obtain the Kitaev honeycomb model. Depending on the choice of $J_\alpha$, this solvable model hosts both a gapped $\mathbb Z_2$ as well as a gapless spin liquid \cite{Kitaev06}. Moreover, starting from the latter (e.g., at the isotropic point $J_x=J_y=J_z$), it is known that adding a small field (i.e., $(3+\sqrt{3}) \nu_\alpha \neq -V_\alpha$) perturbs this into a gapped non-Abelian chiral spin liquid, described by Ising anyon topological order \cite{Kitaev06}.

We note that since we are only considering one particle per triangle, which can moreover not hop off the triangle, the above argument and set-up carries over exactly to the spinless Fermi-Hubbard model.

\subsubsection{Bose-Hubbard model with only on-site interactions \label{subsubsec:BHhopping}}

Here we consider the more restrictive Bose-Hubbard model (i.e., we do not presume any interactions between distinct sites), with $H = H_\textrm{intra} + H_\textrm{inter}$ on the star-honeycomb lattice given by:
\begin{align}
H_\textrm{intra} &= \sum_{j} \left[ \sum_{\alpha=x,y,z} \left( t_c b_{j,\alpha}^\dagger b_{j,c}^{\vphantom \dagger} + t b_{j,\alpha+1}^\dagger b_{j,\alpha}^{\vphantom \dagger} + h.c. \right) + \sum_{a=c,x,y,z} \left(\mu_a n_{j,a} + U n_{j,a} (n_{j,a}-1) \right) \right], \\
H_\textrm{inter} &=  \sum_{\alpha=x,y,z} t_\alpha' \sum_{\langle i,j\rangle_\alpha} \left( b_{i,\alpha}^\dagger b_{j,\alpha}^{\vphantom \dagger} + h.c.\right). \label{eq:intertrianglehop}
\end{align}

Our tactic is to use the hopping term in $H_\textrm{inter}$ to effectively generate the above density-density interactions \eqref{eq:nnrepulsion} by a second-order process.

On upward-pointing triangles (i.e., `A sublattice' of honeycomb lattice), we use the same set-up as above, i.e., we put a single boson with the following parameters:
\begin{align}
t_c &= 2\lambda \sqrt{3\sqrt{3}-5} \approx 0.886 \lambda, \\
t &= \lambda(2-\sqrt{3}+i) = 2\lambda \sqrt{2-\sqrt{3}} e^{i 5\pi/12} \approx 1.035 \; e^{1.309i} \; \lambda,\\
\mu_{\alpha=x,y,z} - \mu_c &= 4 \lambda \left(2-\sqrt{3}\right) + \nu_\alpha \approx 1.072 \lambda + \nu_\alpha, \\
\mu_c &= 2 \lambda (2\sqrt{3}-3) - w \approx 0.928 \lambda  - w.
\end{align}
For now we set $\nu_\alpha = 0$. If we take the limit $\lambda \to + \infty$, then only three states remain at finite energy, namely an effective spin-1/2 qubit at energy $-w$ (per triangle), and an empty triangle with zero energy. The latter will be important when we discuss how the inter-triangle boson hopping induces a diagonal term at second order in perturbation theory. Before we can discuss that, we have to specify the parameters on the other sites.

Let us now consider the downward-pointing triangles (i.e., the other sublattice of the honeycomb lattice). Here we take $U \to +\infty$. (We can similarly take $U \to +\infty$ on upward-pointing triangles, but in that case there is only one boson per triangle, such that it has no effect.) Here we load \emph{three} bosons per triangle (note that our triangles contain \emph{four} sites, including the central site). We choose parameters similar to above, except for taking the complex conjugate, as well as inverting the sign of the chemical potentials:
\begin{align}
t_c &= 2\lambda \sqrt{3\sqrt{3}-5} \approx 0.886 \lambda, \\
t &= \lambda(2-\sqrt{3}-i) = 2\lambda \sqrt{2-\sqrt{3}} e^{-i 5\pi/12} \approx 1.035 \; e^{-1.309i} \; \lambda,\\
\mu_{\alpha=x,y,z} - \mu_c &= - 4 \lambda \left(2-\sqrt{3}\right) - \nu_\alpha \approx -1.072 \lambda - \nu_\alpha, \\
\mu_c &= -2 \lambda (2\sqrt{3}-3) + w \approx -0.928 \lambda  + w.
\end{align}
Again taking $\lambda \to + \infty$, only three finite-energy states remain: firstly, a twofold degenerate `spin-1/2' level with three bosons per triangle, and at an energy cost $w$ above this there is a product state with four bosons.

Since $U \to + \infty$, we have hardcore bosons, and thus effectively we can describe the downward-pointing triangles as containing a single \emph{holon} (rather than three bosons). The holon experiences the opposite chemical potential and magnetic field, i.e., it effectively experiences the same hopping and chemical potential as the boson does on the upward triangles. The net result is just that the inter-triangle hopping in Eq.~\eqref{eq:intertrianglehop} now looks like a pair-creation term. Although this lifts us out of the low-energy space (with an energetic penalty $2w$ which we take to be large), at second-order in perturbation theory it generates a diagonal interaction as in Eq.~\eqref{eq:nnrepulsion}, which is attractive (between the boson on up-triangles and the holon on down-triangles), with $V_\alpha = -\frac{\left(t_\alpha'\right)^2}{2 w}$. Analogous to the previous subsection, we thus \textbf{arrive at an effective spin-1/2 Kitaev model} \eqref{eq:effectiveKitaev}, which is ferromagnetic (if $w>0$), with the dictionary:
\begin{equation}
J_\alpha = - \frac{(2-\sqrt{3})\left(t_\alpha'\right)^2}{12w} \approx -0.0223 \; \frac{\left(t_\alpha'\right)^2}{w} \quad \textrm{and} \quad h_\alpha =
J_\alpha  + \frac{\nu_\alpha}{3+\sqrt{3}} \approx 0.211 \nu_\alpha  -0.0223 \; \frac{\left(t_\alpha'\right)^2}{w}.
\end{equation}

As a brief summary, the hierarchy of energy scales used above is:
\begin{equation}
|\nu_\alpha| \ll |t_\alpha'| \ll |w| \ll \lambda \ll U.
\end{equation}
With this hierarchy, the Bose-Hubbard model (even with only on-site interactions!) can thus realize an exactly-solvable spin liquid. In particular, setting the chemical potential $\nu_\alpha = \frac{\left(t_\alpha'\right)^2}{2(3+\sqrt{3}) w} \approx 0.106 \frac{\left(t_\alpha'\right)^2}{w}$, we obtain the Kitaev honeycomb model (with zero effective magnetic field). To verify that the above algebra does not contain any mistakes, we have explicitly verified (for small systems) that, say, the energy gap of the effective Kitaev model agrees well with what we obtain from exact diagonalization of the original Bose-Hubbard model (for a few random choices of the tuning parameters which respect the above hierarchy).

\section{Details about the Rydberg proposal \label{app:rydberg}}

We consider the setting described in Sec.~\ref{subsec:rydberg} of the main text.

\subsection{Deriving the Hamiltonian}

Let $e_i = 0,1$ ($\tilde e_i=0,1$) denote whether site $i$ is in the $\ket{nS}$ ($\ket{\tilde nS}$) state or not. Finally, $g_i=0,1$ denotes whether the atom is in the ground state; note that $g_i+e_i+\tilde e_i = 1$. In terms of matrices in the basis $\{ \ket{\textrm{gs}}, \ket{nS}, \ket{ \tilde n S } \}$, we can write
\begin{equation}
g = \left( \begin{array}{ccc} 1 & 0 & 0 \\ 0 & 0 & 0 \\ 0 & 0 & 0 \end{array} \right) = \frac{\mathbb I + \mathcal Z + \mathcal Z^\dagger}{3}, \quad
e = \left( \begin{array}{ccc} 0 & 0 & 0 \\ 0 & 1 & 0 \\ 0 & 0 & 0 \end{array} \right) = \frac{\mathbb I + \bar \omega \mathcal Z + \omega \mathcal Z^\dagger}{3}, \quad
\tilde e = \left( \begin{array}{ccc} 0 & 0 & 0 \\ 0 & 0 & 0 \\ 0 & 0 & 1 \end{array} \right)  = \frac{\mathbb I + \omega \mathcal Z + \bar \omega \mathcal Z^\dagger}{3}, \label{eq:RydbergZ}
\end{equation}
where we also expressed them in terms of the diagonal operator $\mathcal Z$ defined in Eq.~\eqref{eq:3statematrices}.

Using the definitions in the main text, the interaction (if $|n-\tilde n| >1$) is well-described by:
\begin{equation}
V = \sum_{\langle i ,j\rangle} \left( U_{nn} e_i e_j + U_{\tilde n \tilde n} \tilde e_i \tilde e_j + U_{n \tilde n} \left[ e_i \tilde e_j + \tilde e_i e_j \right] \right).
\end{equation}
Up to single-site terms, we can use $g_i = 1-e_i-\tilde e_i$ to rewrite the interaction term as
\begin{equation}
V = \sum_{\langle i ,j\rangle} \left( U_{n \tilde n} g_i g_j + \left[ U_{nn} - U_{n\tilde n} \right] e_i e_j + \left[ U_{\tilde n \tilde n} - U_{n\tilde n} \right]\tilde e_i \tilde e_j \right).
\end{equation}
Using Eq.~\eqref{eq:RydbergZ} we can express the two-body term as:
\begin{align}
V &= \frac{1}{9}\sum_{\langle i ,j\rangle} \left( U_{n \tilde n} \left( \mathcal Z_i + \mathcal Z_i^\dagger\right) \left( \mathcal Z_j + \mathcal Z_j^\dagger\right) + \left[ U_{nn} - U_{n\tilde n} \right] \left( \mathcal Z_i + \bar \omega \mathcal Z_i^\dagger\right) \left( \omega\mathcal Z_j + \mathcal Z_j^\dagger\right) + \left[ U_{\tilde n \tilde n} - U_{n\tilde n} \right] \left( \mathcal Z_i + \omega \mathcal Z_i^\dagger\right) \left( \bar \omega\mathcal Z_j + \mathcal Z_j^\dagger\right) \right) \\
&= \frac{1}{9}\sum_{\langle i ,j\rangle} \big( \left[ U_{nn} + U_{\tilde n \tilde n} - U_{n\tilde n}  \right] \mathcal Z_i \mathcal Z_j^\dagger + \underbrace{\left[ \omega U_{nn} + \bar \omega U_{\tilde n \tilde n} + 2 U_{n \tilde n} \right]}_{=2U_{n \tilde n} - \frac{1}{2} \left( U_{nn} + U_{\tilde n \tilde n} \right) + i \frac{\sqrt{3}}{2} \left( U_{nn} - U_{ \tilde n \tilde n } \right) } \mathcal Z_i \mathcal Z_j  \big) + h.c .
\end{align}
Thus far we used $\mathcal Z$ in Eq.~\eqref{eq:3statematrices} with $\phi = 0$. We can set $\phi \neq 0$ to absorb the above complex phase factor. In this case, the interaction term becomes:
\begin{equation}
V = \frac{1}{9}\sum_{\langle i ,j\rangle} \left( \left( U_{nn} + U_{\tilde n \tilde n} - U_{n\tilde n}  \right) \left( \mathcal Z_i \mathcal Z_j^\dagger + \mathcal Z_i^\dagger \mathcal Z_j \right) + \left|\omega U_{nn} + \bar \omega U_{\tilde n \tilde n} + 2 U_{n \tilde n} \right| \; \left( \mathcal Z_i \mathcal Z_j  + \mathcal Z_i^\dagger \mathcal Z_j^\dagger \right) \right) ,
\end{equation}
reproducing Eqs.~\eqref{eq:ZZfromRydberg}, \eqref{eq:Jpm} and \eqref{eq:Jpp}.

\subsection{Experimental values from ARC}

If we want to quantify the interaction strengths for a particular choice of atoms, we need to specify the relevant quantum numbers more carefully. The two Rydberg states correspond to:
\begin{equation}
\ket{nS} \equiv \ket{n,l=0,j=\frac{1}{2},m=\frac{1}{2}} \qquad \textrm{and} \qquad
\ket{\tilde n S} \equiv \ket{n,l=0,j=\frac{1}{2},m=\sigma \frac{1}{2}}.
\end{equation}
Here $\sigma = \pm 1$ is an additional choice; we will discuss its effects shortly. We used the Python package \textsc{arc} \cite{arc} to calculate the (diagonal) interaction strength for particular choices of Alkali atom type, $n$, $\tilde n$ and $\sigma$; in addition, the interaction depends on the azimutal angle $\theta$ between the quantization axis (defining $m$) and the distance vector between the two atoms (i.e., for $\theta = \frac{\pi}{2}$ the atoms are in a plane perpendicular to the quantization axis). More precisely, we calculate the $C_6$ coefficients which determine the Van der Waals interaction $-\frac{C_6}{r^6}$ between two Rydberg states at a distance $r$ (note that with this convention, $C_6>0$ is attractive). I.e., we can equate $U_{nn} = -C_6(nS,nS)/R^6$ where $R$ is the separation between atoms; similarly for $U_{\tilde n \tilde n}$ and $U_{n \tilde n}$.

Let us first consider Potassium ($^{39}$K). For $n=56$, $\tilde n = 58$, $\sigma=1$ and $\theta = \frac{\pi}{2}$, we obtain
\begin{equation}
C_6(nS,nS) = 40.43 \; \textrm{GHz}\left(\mu \textrm{m}\right)^6,
\quad
C_6(\tilde nS, \tilde nS) = 60.81 \; \textrm{GHz}\left(\mu \textrm{m}\right)^6,
\quad
C_6(nS,\tilde nS) = 100.80 \; \textrm{GHz}\left(\mu \textrm{m}\right)^6.
\end{equation}
The fact that $C_6(nS,nS)+C_6(\tilde nS, \tilde nS)- C_6(nS,\tilde nS)$ is very small corresponds to the claim in Sec.~\ref{subsec:rydberg} about the effective Hamiltonian being dominated by a single interaction term. In addition, we observe that the above result is almost independent of the choice of $\sigma = \pm 1$. (Indeed, for $\sigma=-1$, $C_6(nS,nS)$ and $C_6(\tilde nS, \tilde nS)$ do not change, and $C_6(nS,\tilde nS) = 100.38 \; \textrm{GHz}\left(\mu \textrm{m}\right)^6$.) Relatedly, we observe that the results are effectively independent of the azimutal angle $\theta$. The interactions are thus isotropic, which means this can be used to simulate $XY$ magnets in 3D space (see main text). 

We have similarly considered Potassium with $n=89$ and $\tilde n = 92$. Here we also find $C_6(nS,nS)+C_6(\tilde nS, \tilde nS)- C_6(nS,\tilde nS)$ to be very small. More precisely, again setting $\sigma=1$ and $\theta=\frac{\pi}{2}$, we find:
\begin{equation}
C_6(nS,nS) = 8218 \; \textrm{GHz}\left(\mu \textrm{m}\right)^6,
\quad
C_6(\tilde nS, \tilde nS) = 11995 \; \textrm{GHz}\left(\mu \textrm{m}\right)^6,
\quad
C_6(nS,\tilde nS) = 20337 \; \textrm{GHz}\left(\mu \textrm{m}\right)^6.
\end{equation}
We again find that the results are only very weakly dependent on $\sigma$ (and correspondingly $\theta$). For $\sigma=-1$, $C_6(nS,\tilde nS)=20240$ (with the other two values of course unchanged). In terms of $J_{++}$ and $J_{+-}$ in Eqs.~\eqref{eq:Jpm} and \eqref{eq:Jpp}, this means that $|J_{+-}|/J_{++} \approx 0.0009$.

In contrast to Potassium, we find a more significant dependence on $\sigma$ and $\theta$ for Cesium and Rubidium. This comes with advantages and disadvantages. A disadvantage is that it prevents one from using this as a way of simulating the 3D XY model. An advantage is that one can use this dependence to more easily tune such that, e.g., $J_{+-} = 0$. For instance, if we consider Rubidium ($^{85}$Rb) with $n=82$, $\tilde n = 85$, $\sigma=-1$ and $\theta = 0.5066$, we obtain
\begin{equation}
C_6(nS,nS) = 5584.3 \; \textrm{GHz}\left(\mu \textrm{m}\right)^6,
\quad
C_6(\tilde nS, \tilde nS) = 8504.6 \; \textrm{GHz}\left(\mu \textrm{m}\right)^6,
\quad
C_6(nS,\tilde nS) = 14089.2 \; \textrm{GHz}\left(\mu \textrm{m}\right)^6.
\end{equation}
This leads to $|J_{+-}|/J_{++} < 0.00002$. That being said, one does not \emph{have} to use the freedom in the azimutal angle to get a dominant $J_{++}$. E.g., for Rubidium with $n=81$, $\tilde n=84$, $\sigma=-1$ and $\theta = \frac{\pi}{2}$, we have that $J_{+-}$ is less than $4\%$ of $J_{++}$. Or for Cesium ($^{133}$Cs) with $n=84$, $\tilde n = 87$, $\sigma = -1$ and $\theta = \frac{\pi}{2}$, we have $|J_{+-}|/J_{++} \approx 0.01$. 
\end{document}